\documentclass[pre,aps,superscriptaddress,showpacs,floatfix]{revtex4}
\usepackage{amsmath}
\usepackage{amssymb}
\usepackage{amsbsy}
\usepackage{graphicx}
\usepackage{color}

\usepackage{enumerate}
\usepackage{mathtools}
\usepackage{natbib}

\def\be{\begin{equation}}
\def\ee{\end{equation}}
\def\bfi{\begin{figure}}
\def\efi{\end{figure}}
\def\bea{\begin{eqnarray}}
\def\eea{\end{eqnarray}}

\begin{document}

\title{Condensation of Fluctuations in the Ising Model: a Transition without Spontaneous Symmetry
Breaking}  

\author{Annalisa Fierro}
\email{annalisa.fierro70@gmail.com}
\affiliation{CNR-SPIN, c/o Complesso di Monte S. Angelo, via Cinthia - 80126 - Napoli, Italy}

\author{Antonio Coniglio} 
\email{coniglio@na.infn.it}
\affiliation{CNR-SPIN, c/o Complesso di Monte S. Angelo, via Cinthia - 80126 - Napoli, Italy}

\author{Marco Zannetti}
\email{mrc.zannetti@gmail.com}
\affiliation{Dipartimento di Fisica "E. R. Caianiello", 
Universit\`a di Salerno, Via Giovanni Paolo II 132, I-84084 Fisciano (SA), Italy}
\begin{abstract}

  The ferromagnetic transition in the Ising model is the paradigmatic example of
  ergodicity breaking accompanied by symmetry breaking. It is routinely assumed
  that the thermodynamic limit is taken with free or periodic boundary conditions.
  More exotic symmetry-preserving boundary conditions, like cylindrical antiperiodic,
  are less frequently used for special tasks, such as the study of phase coexistence
  or the roughening of an interface.
  Here we show, instead, that when the thermodynamic limit is taken with these
  boundary conditions, a novel type of transition takes place below $T_c$ (the usual Ising transition temperature)
  without breaking neither ergodicity nor symmetry. Then, the low temperature phase is characterized by a regime ({\it condensation}) of strong magnetization's fluctuations which replaces the usual ferromagnetic ordering. This is due to 
critical correlations perduring for all $T$ below $T_c$. The argument is developed
  exactly in the $d=1$ case and numerically in the $d=2$ case.


\end{abstract} 

\maketitle

\section{Introduction}

Our understanding of phase transitions has been shaped to a large extent by the conceptual
structure of the Landau theory~\cite{Landau}, whose key feature is the reduction of symmetry
as the temperature is lowered from above to below the critical point. This is the
process of spontaneous symmetry breaking (SSB), which is manifested through the
appearence of a non vanishing value of an order parameter, such as the magnetization
in a ferromagnet. However, the Landau paradigma does not exhaust the variety of possible
phase transitions. There are transitions which do not involve
SSB, a notable example among these being 
the topological transition in the two-dimensional XY model.
This paper is devoted to the study of another instance of a phase transition
without SSB, which is particularly interesting because it occurs
in the framework of the Ising model where it is generally taken for
granted that the transition ought to take place with the spontaneous breaking of the
up-down symmetry of the interaction.

It is convenient to first recall some well established facts about the connection between
symmetry and ergodicity breaking~\cite{van Enter,Palmer}. To fix the ideas 
consider a $d$-dimensional Ising system on a lattice
of size $V=L^d$ with the energy function (Hamiltonian)
\be
{\cal H}(\mathbf{s}) ={\cal H}_0(\mathbf{s}) + {\cal B}(\mathbf{s}),
\label{Ham.1}
\ee
where $\mathbf{s}=[s_i=\pm 1]$ is a spin configuration,
${\cal H}_0(\mathbf{s}) = -J\sum_{<ij>} s_is_j$ is the nearest neighbours interaction
with ferromagnetic coupling ($J > 0$)
and ${\cal B}(\mathbf{s})$ is the boundary term.
The interaction ${\cal H}_0(\mathbf{s})$ is invariant with respect to spin reversal ($\mathbb{Z}_2$ group).
In the following we shall restrict the boundary conditions to periodic (PBC) and cylindrical
antiperiodic (APBC), both symmetry-preserving.
To briefly recall what these BC prescribe, consider a square lattice. In the PBC
case, spins on opposite edges are coupled ferromagnetically, just like spins in the bulk.
Instead, in the APBC case, spins on one pair of opposite edges
are coupled ferromagnetically, while those on the other pair antiferromagnetically. Hence
\be
{\cal B}^{(p),(a)}(\boldsymbol{s}) = -J \sum_{y=1}^L s_{1,y} s_{L,y} \mp J\sum_{x=1}^L s_{x,1}s_{x,L},
\label{bdr.1}
\ee
where PBC or APBC correspond to the upper or lower sign and $x,y$ denote horizontal
and vertical directions.  In the following we shall use the $(p)$ and $(a)$ superscripts for PBC and APBC, respectively, except when not required by clarity.
The reason for considering these two boundary conditions is to
show that while with PBC the usual ferromagnetic transition involving ergodicity and symmetry
breaking takes place, in the APBC case we are presented with the novel and
qualitatively different scenario of a transition {\it without} ergodicity and symmetry breaking,
whose low temperature phase is {\it critical} all the way down to $T=0$.

\begin{figure}[ht]
\centering
\includegraphics[width=5cm]{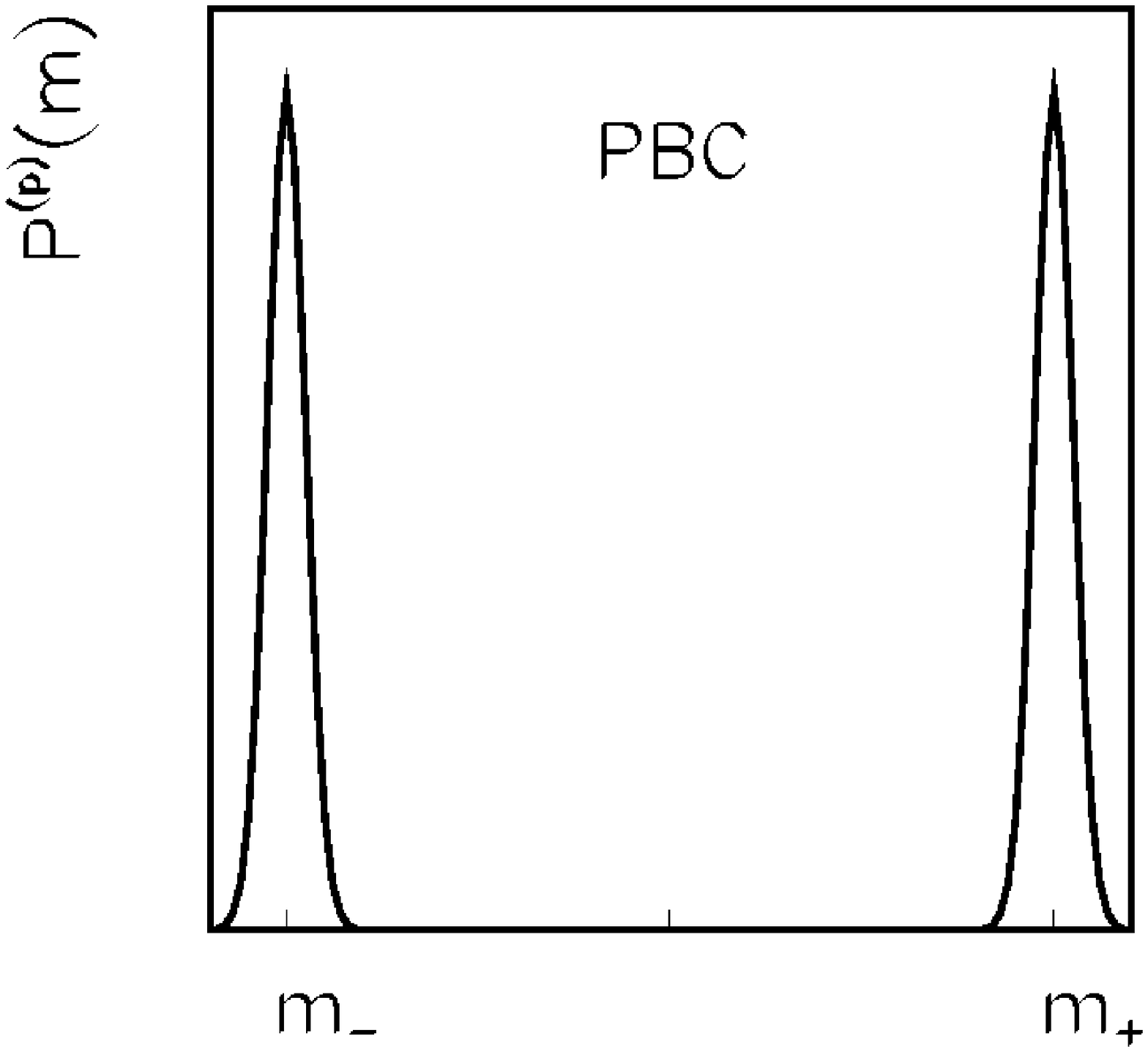}
\includegraphics[width=5cm]{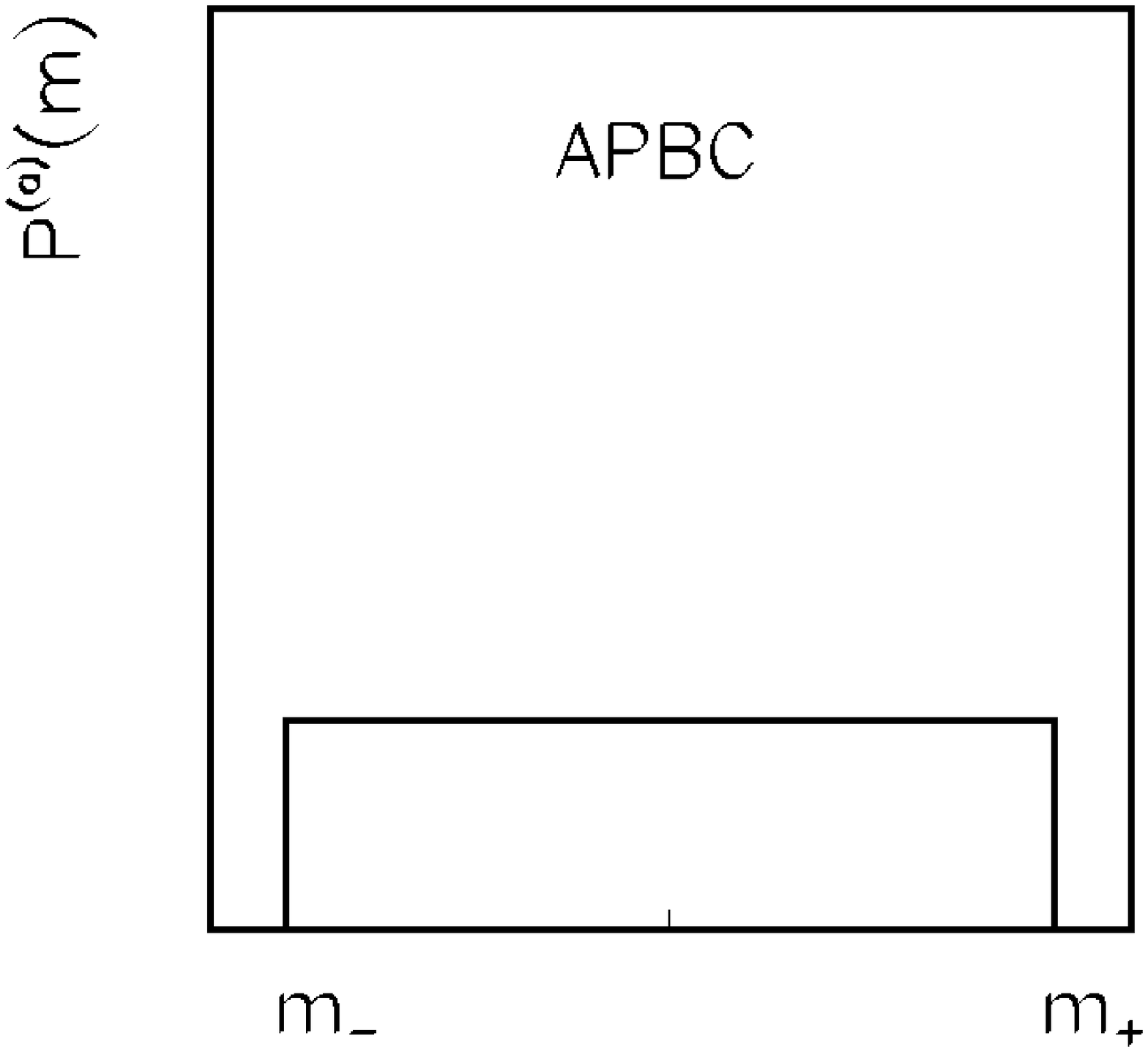}
\caption{Schematic magnetization distributions at some generic $T$ below $T_c$: to the left with PBC and to the right with APBC.} 
\label{fig:1}
\end{figure}

As time evolves the microscopic state executes a trajectory $[\mathbf{s}(t)]$ inside the
phase space of all possible configurations $\Omega = [\mathbf{s}]$.
In general, if $V$ is finite and $T > 0$ the system is ergodic, independently of the BC choice.
This means that all microstates in $\Omega$ are dynamically accesssible from any one of them.
In thermal equilibrium
the trajectory samples phase space according to the time-invariant
Boltzmann-Gibbs distribution
\be
P(\mathbf{s}|\mu) = \frac{e^{-\beta {\cal H}(\mathbf{s})}}
{Z(\mu)},
\label{Ham.2}
\ee
where $\mu$ collects the state parameters $V,T$ and the boundary condition.
However, in the thermodynamic limit ($V \to \infty$) and for sufficiently low $T$,
ergodicity may fail, depending on BC. In fact, 
this is what happens with PBC. By lowering $T$ below the critical
temperature $T_c$ phase space breaks up into the two dynamically
disjoint components $\Omega_\pm$ of the states with positive and negative magnetization,
which transform one into the other under inversion ($\mathbb{I}\Omega_\pm = \Omega_\mp$).
Ergodicity is globally broken because the trajectory remains confined
within the same component in which it has originated, but continues to hold
separately within each component. Then, individual trajectories
do not anymore sample $\Omega$ according to the Boltzman-Gibbs distribution, but
according to non-symmetric distributions $P_\pm(\mathbf{s})$, with support over $\Omega_\pm$
and related by $P_\pm(\mathbb{I} \mathbf{s}) = P_\mp(\mathbf{s})$. These
distributions correspond to the two possible ferromagnetic pure states formed
below $T_c$, while the symmetric Boltzman-Gibbs distribution~(\ref{Ham.2}) becomes the even mixture
of these. The corresponding magnetization density probability distribution takes the double peak form
  \be
  P^{(p)}(m) = \frac{1}{2}[P_+(m) + P_-(m)],
  \label{ens.1}
  \ee
as schematically depicted in the left panel of Fig.\ref{fig:1}. The peaks are centered about
the spontaneous magnetization values $m_\pm$.

Clearly, since in a single experiment the trajectory is confined inside
either one of $\Omega_\pm$, only one peak at the time can be observed.
The inter-component fluctuations~\cite{Palmer} connecting one peak to the other
  are not physical. The even
  form~(\ref{ens.1}) of the distribution can be observed only
  by carrying out the experiment on an ensemble of identically prepared
  systems. In that case, the ensemble average of the magnetization vanishes and the information
  on the spontaneous magnetization is recovered from the second moment
  \be
  \langle m^2 \rangle^{(p)} = m^2_\pm.
  \label{ens.3}
  \ee
 In the effort to make precise the
concept of spontaneous magnetization, Griffiths~\cite{Griffiths}
proved that, in the absence of an external field explicitly breaking the symmetry,
the probability of finding a value of the magnetization outside the interval
$[m_-,m_+]$ vanishes in the thermodynamic limit, without being able, though, to
pinpoint the shape of the distribution within the interval. Nonetheless, as a corollary there follows the
inequality
\be
\langle m^2 \rangle \leq m_\pm^2.
\label{Griffiths.1}
\ee
For future reference we mention here that ~Eq.(\ref{ens.1})  saturates Griffiths
  inequality as an equality. 
  
  The one above outlined is the standard SSB picture, which is radically subverted if the
  thermodynamic limit is taken imposing APBC. The main purpose of the present paper is to show that,
  with this choice of BC and in the thermodynamic limit,
  below the standard critical point at $T_c$ there is a novel low $T$ phase whose prominent features, as
  stated above, are i) the absence of SSB and ii) the presence of long range correlations.
 APBC have been usually
employed as a mean to artificially create an interface in order to study details of phase coexistence, like surface tension~\cite{Gallavotti,Delfino},  
or to focus on the structure of the interface itself and its fluctuations, with particular interest
in the roughening transition \cite{Hasenbusch}.
Here we broaden the scope, analysing the system's global behavior as temperature is varied. In the studies quoted above the interest was essentially on the structure of the interface and its fluctuations, without considering the translations of the interface.
Here, instead, we are interested in the fluctuations responsible of the displacement of the interface as a whole in the direction along which APBC are imposed, which sustain both ergodicity and criticality.

For the $d=1$ system
we show, on the basis of exact results, that the new phase, characterized by the uniformity of the magnetization distribution
in the interval $[m_-,m_+]$ as sketched in the right panel of Fig.\ref{fig:1}, is formed at $T=0$. The reason for this is that ergodicity is not broken. 
        The trajectory tipically visits
        the subset of configurations characterized by one domain wall separating two oppositely
        ordered large domains. These configurations are dynamically connected and since
        the wall can freely wander through the system,
        all values of the magnetization
        in the interval $[m_-,m_+]$ become equally probable.
The fluctuations
spanning this interval are 
of the intra-component kind~\cite{Palmer}, implying that
the null result $\langle m \rangle^{(a)} = 0$ is physical, i.e. the absence of spontaneous magnetization
is obtained from the time average
on a single experiment. In this case ensemble averages and time averages do coincide.
This transition without SSB is revealed by the second moment
which, as a consequence of the uniformity of probability, goes like
\be
\langle m^2 \rangle^{(a)} = \left \{ \begin{array}{ll}
0, \quad T > 0, \\
  \frac{1}{3} m^2_\pm,\quad T = 0,    
        \end{array}
        \right .
        \label{ens.5}
        \ee
        and satisfies Griffiths inequality strictly.
The significant difference with Eq.~(\ref{ens.3}) is that now the finite value of
  $\langle m^2 \rangle$ does not originate from ordering of the system,
but from macroscopic fluctuations of the magnetization due to correlations extending
over the entire volume of the system (hence critical).
We shall refer to such a transition as one of
{\it condensation of fluctuations}, as opposed to the usual ferromagnetic transition.
The gross features of the above picture are numerically confirmed with good precision
in the $d=2$ case for $0 < T < T_c$, where $T_c$ is the Ising critical temperature.
The $T=0$ state is excluded due to the pinning of the interface when the straight geometry
is reached.

The paper is organized as follows:  the case of the $d=1$ model is presented in
Section~\ref{sec2}, where the $T=0$ transition is 
analysed in detail through exact results. The $d=2$ case is investigated numerically in
Section~\ref{sec4}, where the scenario is expanded and enriched by the finite $T_c$ value.
Conclusions and the outlook are presented in Section~\ref{sec5}.

\section{Ising model $d=1$}
\label{sec2}

The whole structure outlined in the Introduction can be neatly illustrated in 
the exactly soluble one-dimensional model.
Consider an Ising chain of length $L$ with energy function
\be
{\cal H}_L(\boldsymbol{s}) = -J\sum_{i=1}^{L-1} s_is_{i+1} + {\cal B}(\mathbf{s}),
\label{Is.1}
\ee
where the boundary term reads
\be
{\cal B}(\mathbf{s}) =   \left \{ \begin{array}{ll}
         -Js_Ls_1,\;\; $PBC$, \\
         Js_Ls_1,  \;\; $APBC$.     
        \end{array}
        \right .
        \label{BCS.1}
        \ee

The magnetization probability distribution has been computed exactly
        for arbitrary $L,T$ and various BC in Ref.~\cite{Antal}. In the thermodynamic limit
        and $T > 0$ one finds $P(m) = \delta(m)$ in all cases.
        But, the dependence on BC emerges at $T = 0$ yielding the double peak characteristic of SSB
        with PBC
\be
        P^{(p)}(m) = \frac{1}{2} [\delta(m+1) + \delta(m-1)],
        \label{zfpb.1}
        \ee
        as opposed to the uniform shape with APBC 
        
        \be
        P^{(a)}(m) =   \left \{ \begin{array}{ll}
        1/2 ,\;\; m \in [-1,1], \\
         0,  \;\; m \notin [-1,1],     
        \end{array}
        \right .
        \label{zfapb.1}
        \ee
        which correspond to the distributions of Fig.\ref{fig:1} with $|m_\pm|=1$.
        Notice that, although different, both distributions comply with the Griffiths
        theorem previously quoted.
        As anticipated in the Introduction, the difference in the ergodic properties
        is at the root of the difference in shape of the probability distributions.
When PBC are imposed the ground state is degenerate and 
        it is given by either
        one of the two ordered configurations, with all spins up $\mathbf{s}_+=[s_i=+1]$
        or all spins down $\mathbf{s}_-=[s_i=-1]$. These are dynamically disconnected
        since the switch from one to the other would require activated moves.
        In other words, at $T=0$ ergodicity is broken and $\mathbf{s}_\pm$ coincide with the two
        absolutely confining ergodic components $\Omega_\pm$. Consequently,
        $P^{(p)}(m)$ is the \emph{mixture} obtained by evenly mixing the two pure
        states $P_\pm(m)=\delta(m \mp 1)$, which means, as explained in the Introduction, that
in spite of the parity of $P^{(p)}(m)$
        the system orders and SSB takes place.

        Conversely,
        in the APBC case configurations of lowest energy are those with one defect~\cite{Antal}, or
        domain wall, which are dynamically connected since the defect can freely travel
        along the system by performing random walk at no energy cost. Ergodicity is not broken and
        all values of $m$ can be sampled even in a single experiment.
        Thus, the uniformity of $P^{(a)}(m)$ signifies absence of ordering and of SSB.
        This type of transition, consisting in the appearence of macroscopic fluctuations of $m$
        and without ordering is the condensation of fluctuations.

Since in both cases the symmetry of $P(m)$ is preserved at all temperatures, the two transitions
are revealed by the
second moment $\langle m^2 \rangle$, which jumps from zero at $T > 0$  to
the $T=0$ finite values
\be
\langle m^2 \rangle  =  \left \{ \begin{array}{ll}
         1,\;\; $PBC$, \\
         1/3,  \;\; $APBC$.
        \end{array}
        \right .
        \label{fbc.13}
        \ee
        In the first case $\langle m^2 \rangle$ 
        contains the information on the spontaneous magnetization,
        while in the second one expresses
        only the size of fluctuations.

        \subsection{Correlation function}

\begin{figure}[ht]
\centering
\includegraphics[width=5cm]{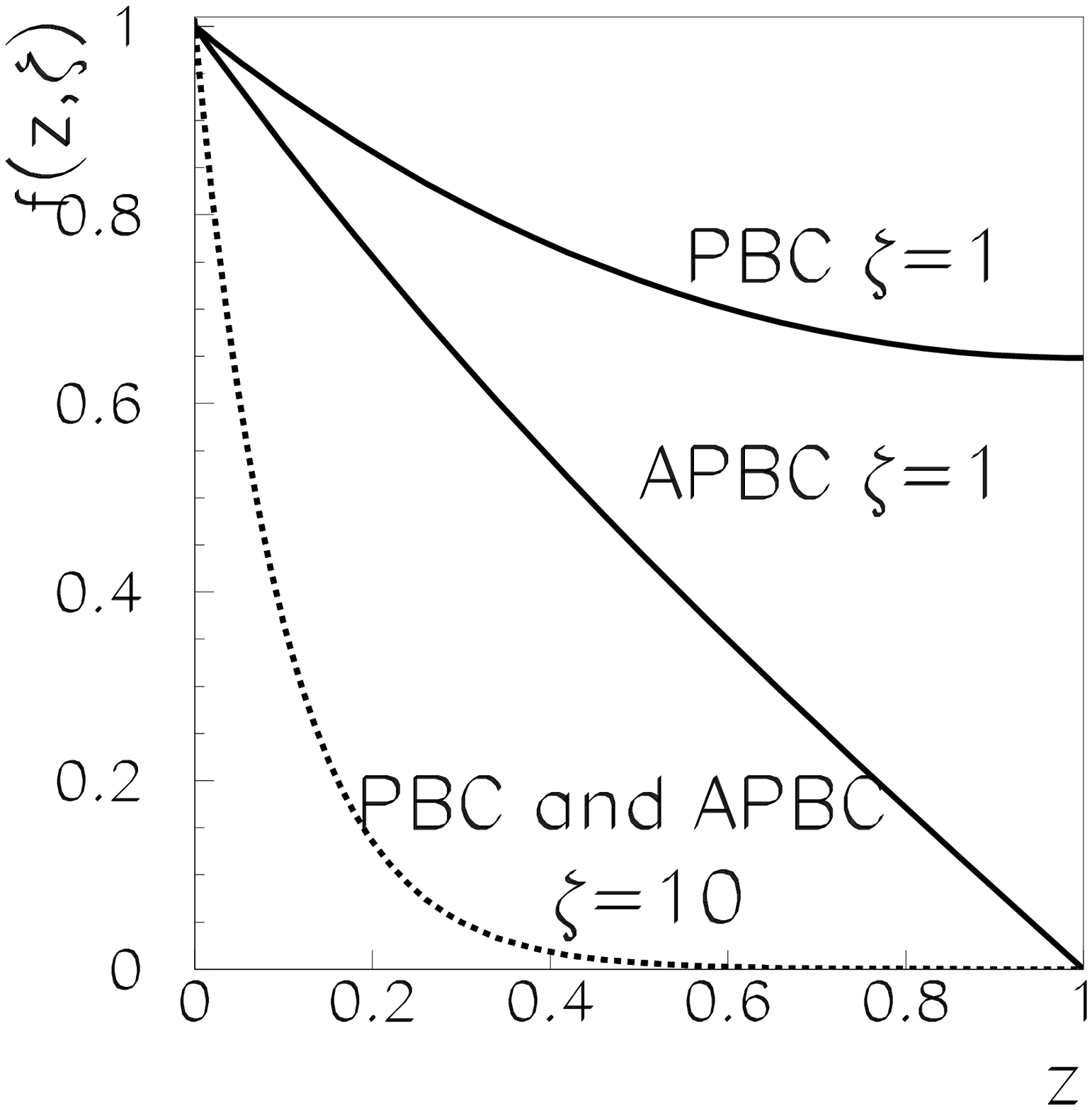}
\includegraphics[width=5cm]{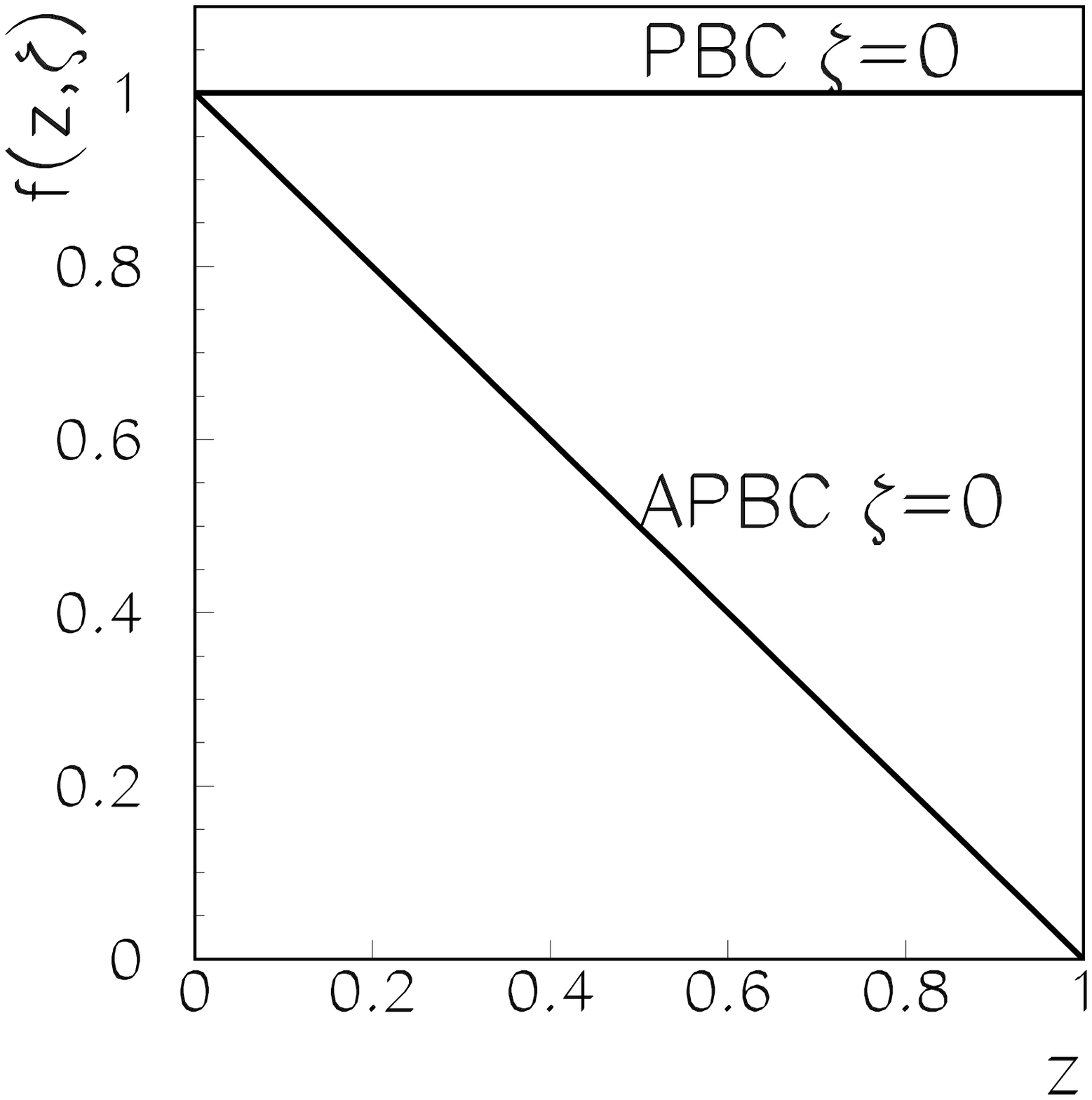}
\caption{Scaling functions from Eqs. (\ref{scal.2}) and (\ref{scal.2bis}), in the $1d$ Ising model with PBC and APBC.
Left panel corresponds to finite temperature and right panel to $T=0$.
}
\label{fig:2}
\end{figure}

        Deeper insight into the difference between the ordering and the condensation transition
        is gained from the correlation function. Using again the general results of Ref.~\cite{Antal},
        the correlation function obeys the scaling form
        \be
        C(r,T,L) = \langle s_is_j \rangle - \langle s_i \rangle \langle s_j \rangle = f(z,\zeta),
        \label{scal.1}
        \ee
        where 
        \be
        r = |i-j| \leq L , \,\,\,\, z=\frac{r}{L/2}, \,\,\,\, \zeta= \frac{L}{2\xi},
        \label{zeta.1}
        \ee
        and $\xi = -[\ln \tanh(J/T)]^{-1}$ is the correlation length in the infinite system.
        Since the scaling functions
 \be
f^{(p)}(z,\zeta) =   
\frac{\cosh \left (\zeta(1-z) \right )}{\cosh(\zeta)},
\label{scal.2}
\ee
\be
f^{(a)}(z,\zeta) =   
\frac{\sinh \left (\zeta(1-z) \right )}{\sinh(\zeta)},
\label{scal.2bis}
\ee
are forced by the BC to be even or odd under space reversal
        \be
f^{(p)}(z,\zeta) = f^{(p)}(z^{\prime},\zeta), \quad f^{(a)}(z,\zeta) = -f^{(a)}(z^{\prime},\zeta),  \quad z^{\prime}=2-z,
\label{scal.3}
        \ee
        we shall consider the behavior only in the first half of the interval $z \leq 1$,
        no new information on the correlations being obtained from the second half.
        The BC dependent finite size effects are negligible if $\xi \ll L$, i.e. $\zeta \gg 1$,
        while do play a role in the opposite regime $\xi \gtrsim L$, i.e. $\zeta \lesssim 1$.
        In the high $T$ regime with $\zeta \gg 1$ exponential decay, independent of
        BC, is found 
        \be
        f(z,\zeta) = e^{-\zeta z} + O(1/\zeta),
        \label{scal.4}
        \ee
        as shown by the dotted curve in Fig.\ref{fig:2}.
        If $T$ is lowered in the $\zeta \lesssim 1$ region, BC dependent
        finite size effects become detectable, driving the slower PBC decay 
        above the APBC one. Finally, at $T=0$, that is for $\zeta = 0$,
        from Eq.~(\ref{scal.2}) follows
\be
\lim_{\zeta \to 0} f(z,\zeta) =  \widehat{f}(z)   = \left \{ \begin{array}{ll}
         1,\;\; $PBC$, \\
         1 - z,  \;\; $APBC$,   
        \end{array}
        \right .
        \label{scal.5}
        \ee
        which shows that the PBC correlation function does not decay, irrespective of the size of $L$,
while in the APBC one there remains an $L$-dependent decay (right panel of Fig.\ref{fig:2}).

        It is important to realize that this difference of behavior is the direct
        consequence of the different distributions and consequently of the different ergodic properties previously discussed.
        To the mixed state~(\ref{zfpb.1}) in configuration space there corresponds
        the mixture of the two pure states concentrated on $\mathbf{s}_\pm$
        \be
        P^{(p)}(\mathbf{s}) = \frac{1}{2} \left [  \delta(\mathbf{s} - \mathbf{s}_+)
         + \delta(\mathbf{s} - \mathbf{s}_-)  \right ],
        \label{ground.1}
        \ee
        from which immediately follows the above result in the first line of Eq.~(\ref{scal.5}) for PBC, since $\langle s_is_j \rangle = 1$ for
        any $r$ and $\langle s_i \rangle = \langle s_j \rangle = 0$. In other words, the correlation function is no-clustering as a consequence
        of ergodicity breaking. Instead, with APBC matters are quite different. As explained previously,
        due to the twisted BC, the lowest energy configurations contain one defect.
        Therefore, there is probability $r/L$ that the two sites $i$ and $j$
        are on opposite sides of the defect and probability $1-r/L$ for them to be on the same side,
        from which follows the second line of Eq.~(\ref{scal.5}).
Hence, at $T=0$ the state is \emph{critical}, since the correlation length is of the order of the size of
the system, which accounts for the macroscopic
fluctuations of $m$, as shown by Eq.~(\ref{fbc.13}).
This $L$ dependence
generates the sharp distinction between the short and large distance behavior,
depending on the scale of $r$, when $L \to \infty$.
If $r$ is kept fixed, then $z \to 0$ as $L \to \infty$ and
\be
\lim_{L \to \infty} \widehat{f}^{(a)}(z) = 1,
\label{p.6}
\ee
while, if $z$ is kept fixed
\be
\lim_{L \to \infty} \widehat{f}^{(a)}(z) = 1-z.
\label{p.7}
\ee
Thus, ergodicity looks broken at short distance ($r \ll L$), while
on the scale of $L$, no matter how large $L$ is taken, there will be always a defect
going by, causing decorrelation and restoring the clustering property.
Borrowing terminology from aging systems,
this is an instance of {\it weak ergodicity breaking}~\cite{BCKM,MZ},
which in the end means that ergodicity is not broken.

In closing this section, we point out that from the comparison of Eqs.~(\ref{scal.2})
or~(\ref{scal.2bis}) with the generic finite size scaling form of the correlation function
\be
C(r,T,L)=\frac{1}{r^{2(d-D)}}f(z,\zeta),
\label{fsz.1}
\ee
where $D=(d+2-\eta)/2$ is the fractal dimensionality of correlated clusters \cite{correlated-percolation},
there follows that in $d=1$ the correlated clusters are compact, i.e. $D=1$ implying
$\eta=1$.

\section{Ising model $d=2$}
\label{sec4}

Let us now consider a two-dimensional finite square lattice containing
$V=L^2$ sites.
The BC interaction term~(\ref{bdr.1}) has been discussed in the Introduction.
The exact result for the spontaneous magnetization of the infinite system is given by \cite{McCoy} 
\be
m_{\pm}  =   \left \{ \begin{array}{ll}
         0,\;\; $for$ \quad T \geq T_c, \\
         \pm [1-\sinh^{-4}(2J/T)]^{1/8},  \;\; $for$ \quad T < T_c,     
        \end{array}
        \right .
        \label{SSB.01}
        \ee
        where $T_c = 2.269 J$ is the critical temperature. From now on we shall set $J=1$.

\begin{figure}[ht]
\centering
\includegraphics[width=5cm]{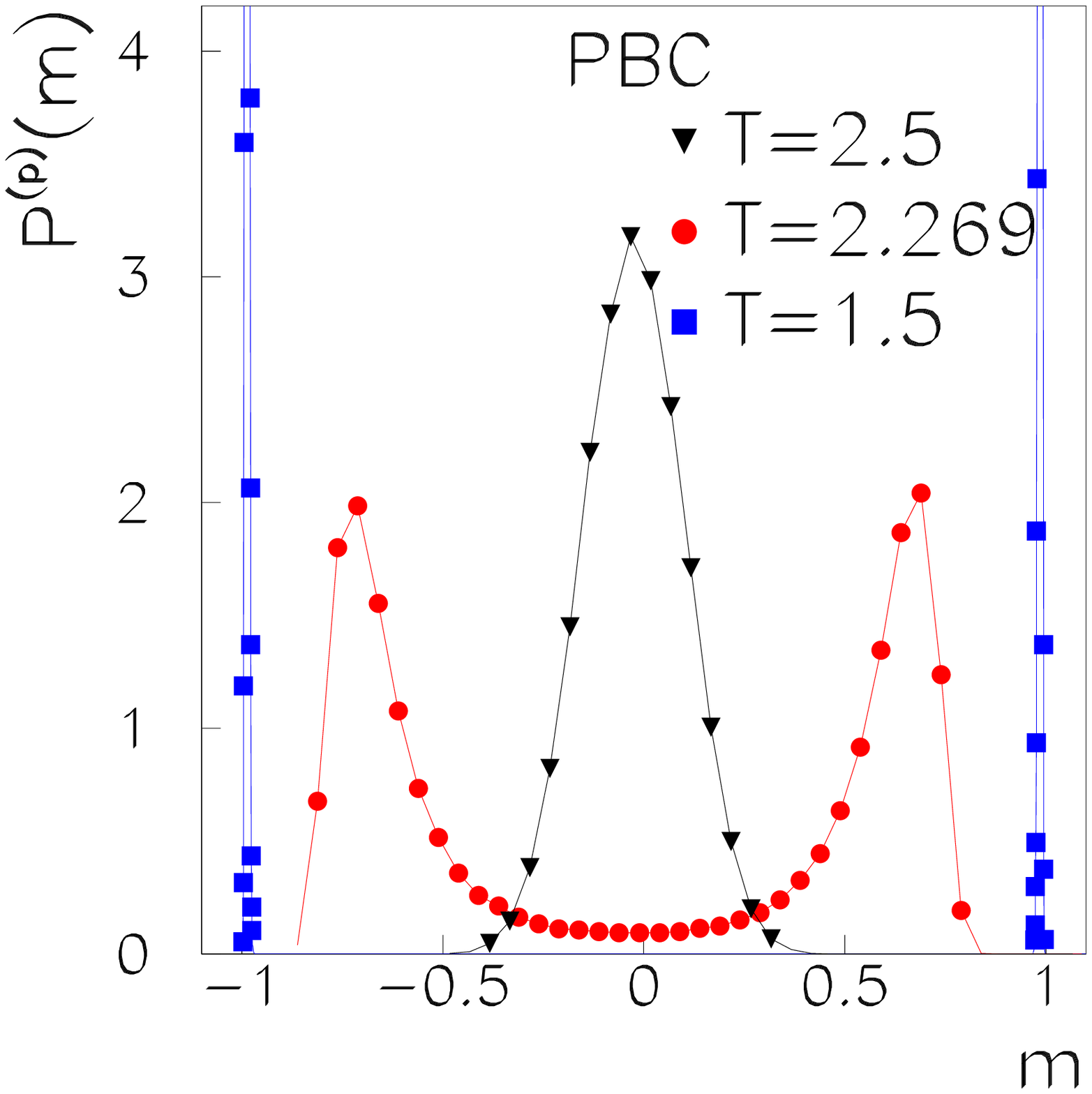}
\includegraphics[width=5cm]{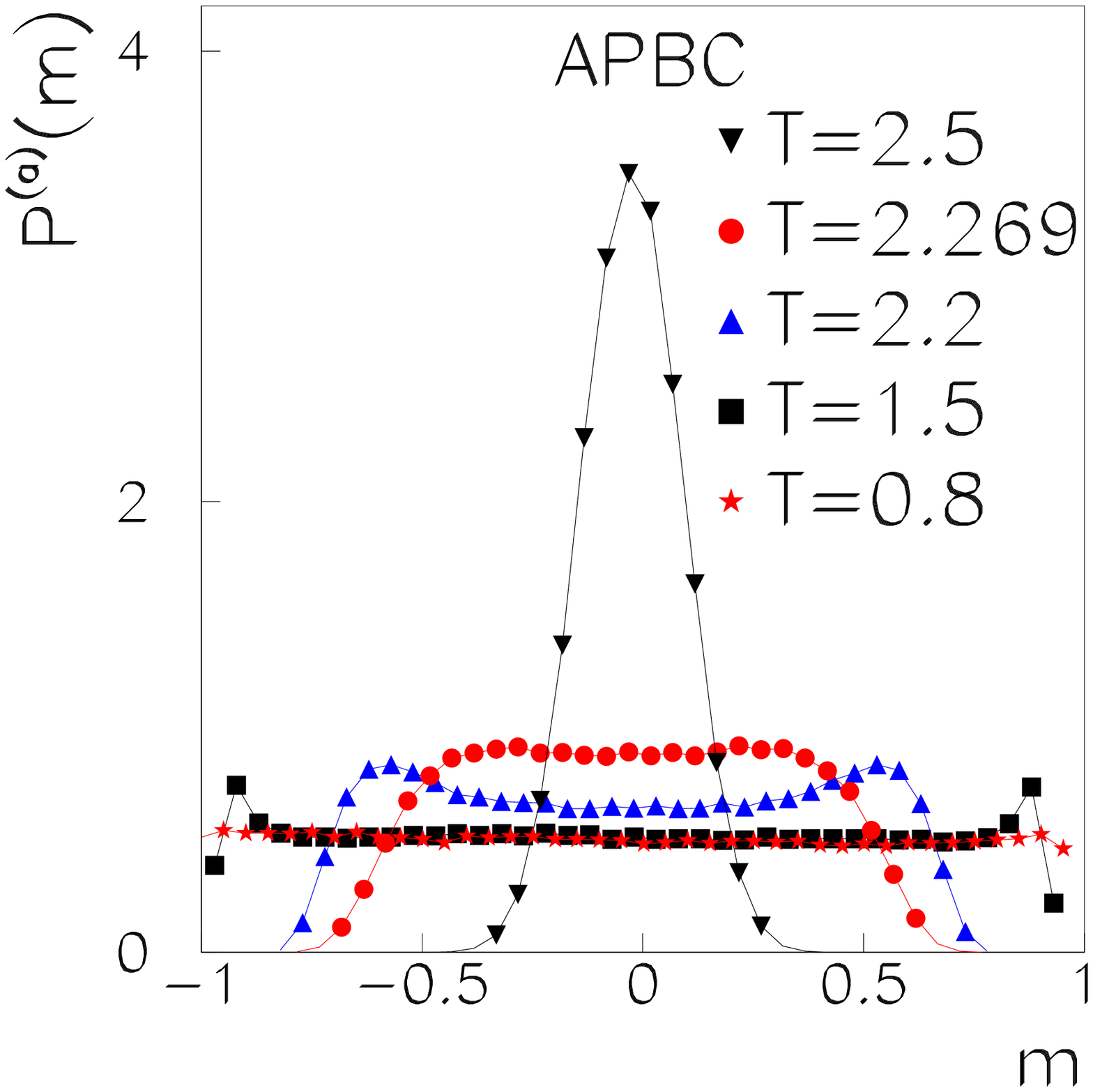}
\includegraphics[width=5cm]{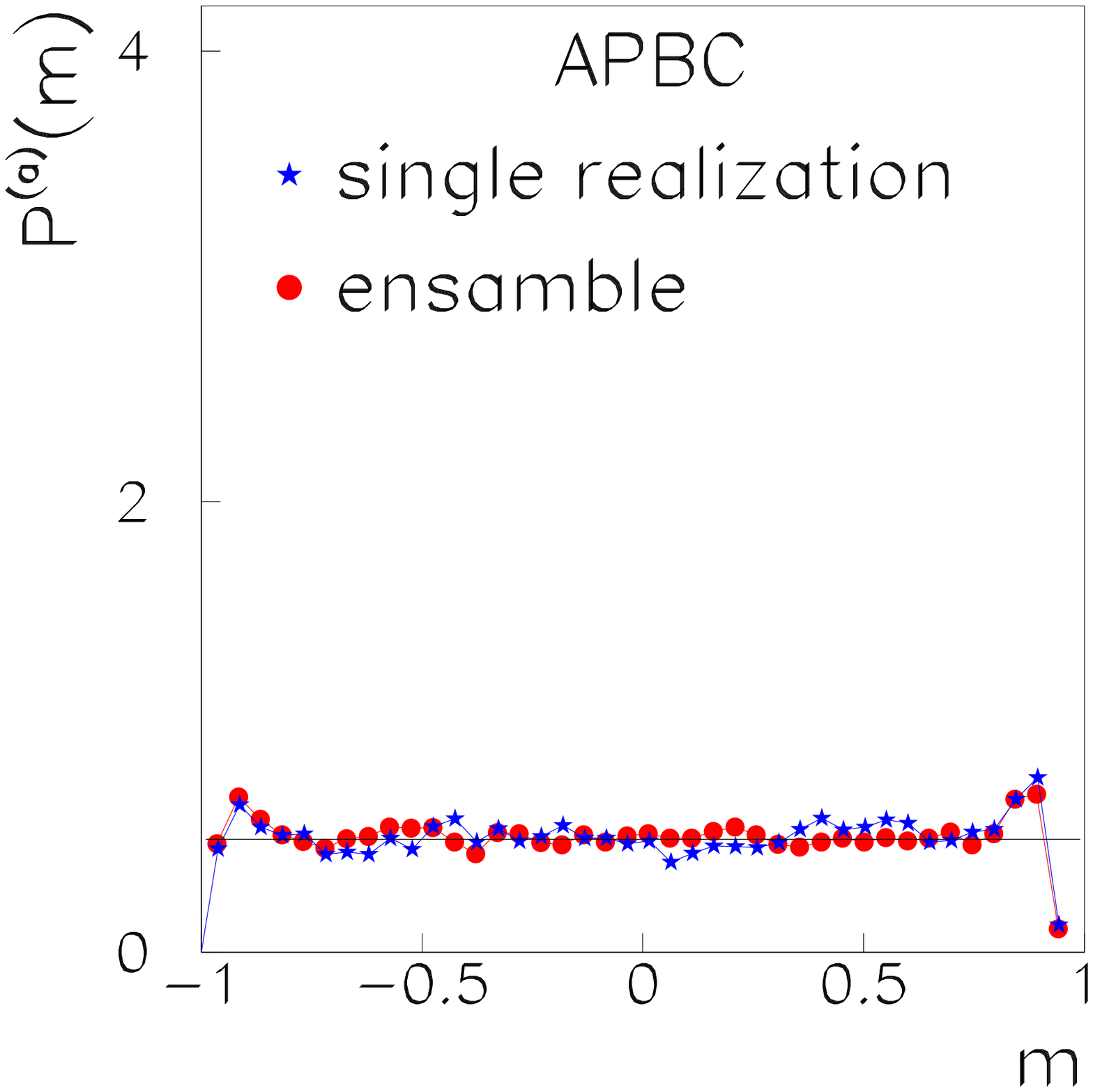}
\caption{Magnetization distribution in the 2d Ising model with PBC 
(left) and APBC (center),  for different $T$ and $L=64$. 
Continuous lines are guides to the eye. Right panel: Comparison of $P^{(a)}(m)$ from a single realization (stars) and
from an ensemble of $1.9 \times 10^4$ independent configurations (circles)
for $T=1.5$ and $L=64$. 
}
\label{fig:3}
\end{figure}

We have numerically extracted $P(m)$, above and below $T_c$, by preparing an $L=64$ system at $T=0$ and letting it to thermalize at the final temperature $T$. We have
sampled $m$ every $1000$ Monte Carlo steps from $50$ independent realizations of total length $10^7$ Monte Carlo steps. The outcome is displayed in the left (PBC) and center (APBC) panels of  Fig.\ref{fig:3}.
The overall pattern replicates the structure observed in the
$d=1$ case.
Above $T_c$ the distribution
is independent of BC. In both cases there is a peak centered on the origin, which is expected to narrow
toward  $\delta(m)$ as $L$ grows. Below $T_c$, instead, the dependence on BC is strong.
In the left panel there appears the growth with decreasing temperature  of the bimodal distribution, characteristic of ergodicity breaking and SSB,
with the two peaks centered about $m_-$
and $m_+$~\cite{Bruce,Binder}.     
Conversely, in the center panel as $T$ goes below $T_c$ the distribution spreads out over the interval $[m_-,m_+]$. For $T$ sufficiently low the data show convergence  toward a limit distribution which is uniform to a good approximation. Ergodicity and symmetry are preserved as it is illustrated in right panel of 
Fig.\ref{fig:3},
where the distribution of $m$ computed at $T=1.5$ from a single realization of $10^7$ Monte Carlo steps is compared with the one extracted from an ensemble of $1.9 \times 10^4$ independent configurations. A similar computation carried out in the PBC case (not displayed here) shows only one peak  from the single time series, as opposed to the two peaks obtained from the ensemble. 
 
The comparison of $\langle m^2 \rangle^{(p)}$ with
\begin{figure}[ht]
\centering
\includegraphics[width=7.5cm]{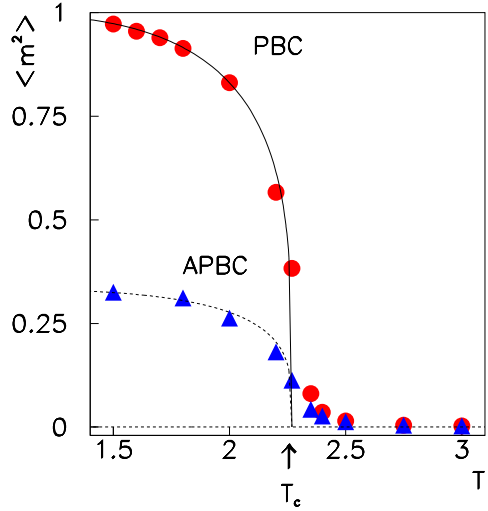}
\caption{Temperature dependence of 
$\langle m^2 \rangle$ in the $2d$ Ising model with $L=64$. PBC (circles) and APBC
(triangles).
The continuous line and the dashed line are the plots of $m^2_{\pm}$ and $\frac{1}{3}m^2_{\pm}$ computed from Eq. (\ref{SSB.01}).}
\label{fig:4}
\end{figure}
$m^2_{\pm}$ and of $\langle m^2 \rangle^{(a)}$ with
$\frac{1}{3}m^2_{\pm}$ is displayed in Fig.\ref{fig:4},
showing that Eqs.~(\ref{ens.3}) and~(\ref{ens.5}) are verified with good precision.

\subsection{Correlation function}

After checking that $\langle s_i \rangle = 0$,  the correlation functions are computed along
the $x$-horizontal and $y$-vertical directions in the following way:
\begin{equation}
C_x(r,L,T)=\frac{1}{L(L-r)}\sum_{y=1}^L \sum_{ij}  \langle s_is_j\rangle,
\end{equation}
\begin{equation}
C_y(r,L,T)=\frac{1}{L(L-r)}\sum_{x=1}^L \sum_{ij}  \langle s_is_j\rangle,
\end{equation}
where the inner sums are over $i$ and $j$ such that $y_i=y_j=y$, $|x_i-x_j|=r$ in the first one
and $x_i=x_j=x$, $|y_i-y_j|=r$ in the second one.
$L-r$ is the number of pairs $(i,j)$ with the same $y$, or $x$, and at a distance $r$.
The average is taken over a set of $1000$ independent realizations. 

In the PBC  case there is  isotropy between the $x$ and $y$ directions, with
$C^{(p)}_x(r,L,T)=C^{(p)}_y(r,L,T) = C^{(p)}(r,L,T)$ symmetric 
under space reversal $r \to r^{\prime} =L-r$, while in the APBC case there is anisotropy, since
$C^{(a)}_x(r,L,T)$ is symmetric and $C^{(a)}_y(r,L,T)$ is antisymmetric.
As in the $d=1$ case, we will restrict the study of these functions to the
half interval $r \in [0,L/2]$.

\subsection{PBC}

We have plotted $C^{(p)}(r,L,T)$ for $T$ above, at and below $T_c$ in Fig.\ref{fig:5} and Fig.\ref{fig:6}:

\begin{itemize}

\item \underline{$T > T_c$} -  Finite size scaling holds in the form of Eq.~(\ref{fsz.1}) with
  $D=1.875$ since $\eta=1/4$. Here and in the following we shall keep using the
  scaling variables $z$ and $\zeta$ defined in Eq. (\ref{zeta.1}).
As in the $d=1$ case, at a temperature $T$ sufficiently higher than $T_c$, such that
$\zeta \gg 1$, the correlation function becomes independent of $L$ and decays exponentially
to zero like $e^{-r/\xi}$. This is shown in the left panel of Fig.\ref{fig:5}, where
$r^{0.25}C^{(p)}(r,L,T)$ has been plotted against $r/\xi$ for different $T$ and $L$,
after extracting $\xi$ as a linear fit parameter from the semilog plot.
\begin{figure}[ht]
\centering
\includegraphics[width=5cm]{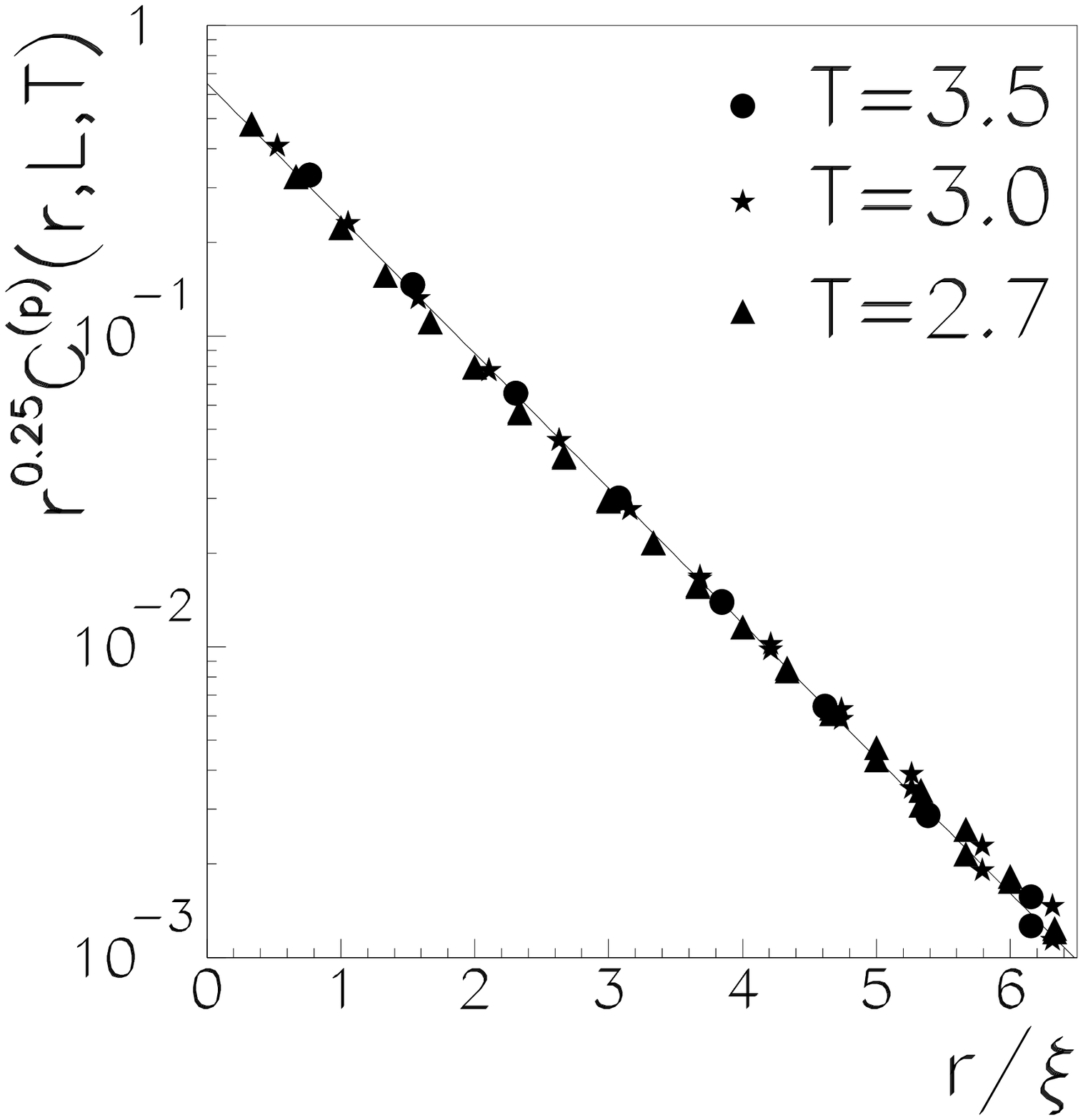}
\includegraphics[width=5cm]{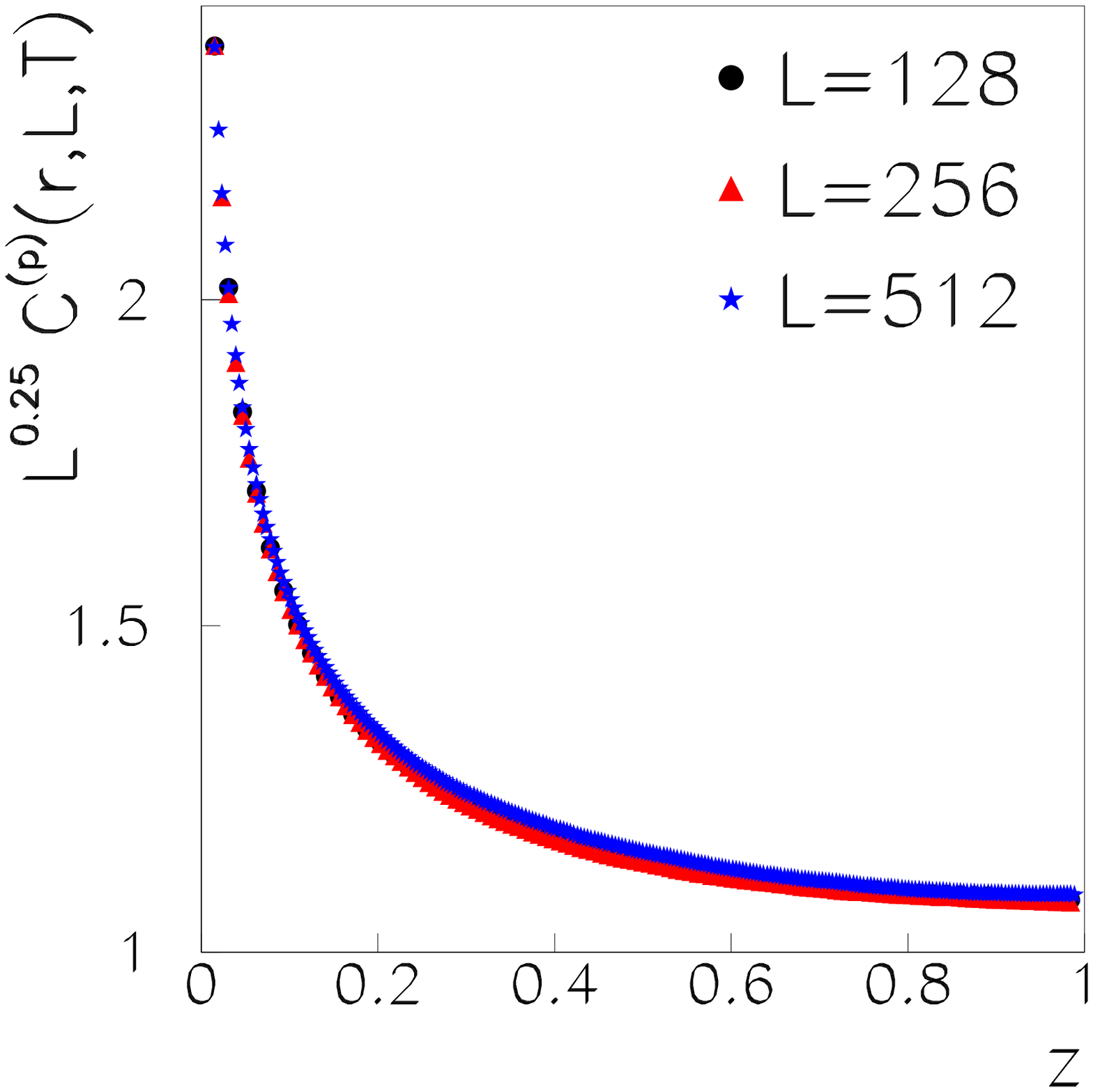}
\caption{Left panel: Plot of $r^{0.25}C^{(p)}(r,L,T)$ against $r/\xi$
in the $2d$ Ising model with PBC. Independence from $L$ is illustrated by the superposition of the data taken with $L=512$ and $L=256$ on the continuous line representing $e^{-r/\xi}$.  
Right panel: data collapse at criticality with PBC. }
\label{fig:5}
\end{figure}

\item \underline{$T = T_c$} - Since $\xi = \infty$, the finite size scaling form reduces to
\be
C^{(p)}(r,L,T_c) = \frac{1}{r^{1/4}}f_c^{(p)}(z),
\label{sim2d.3bis}
\ee  
as demonstrated in right panel of Fig.\ref{fig:5}, where the data taken for different system's sizes
have been collapsed by plotting $L^{1/4} C^{(p)}(r,L,T_c)$ against $z=2r/L$.

\item \underline{$T < T_c$} -  Below $T_c$ the correlation function decays rapidly to a flat
  plateau  (left panel of Fig.\ref{fig:6}), whose height $q$ increases by lowering the temperature according to
  $q = m^2_\pm$, as demonstrated in right panel.
\begin{figure}[ht]
\centering
\includegraphics[width=5cm]{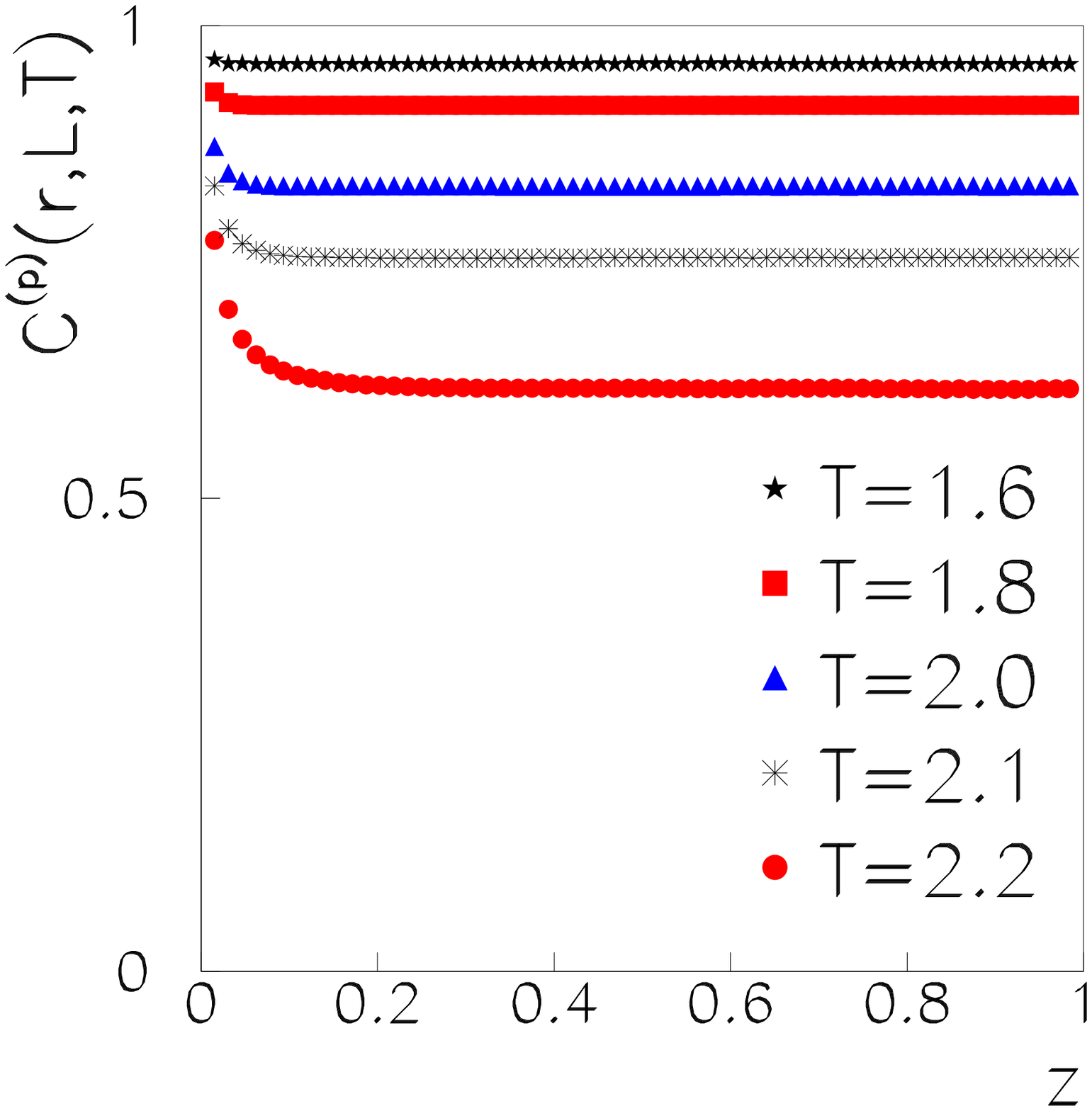}
\includegraphics[width=5cm]{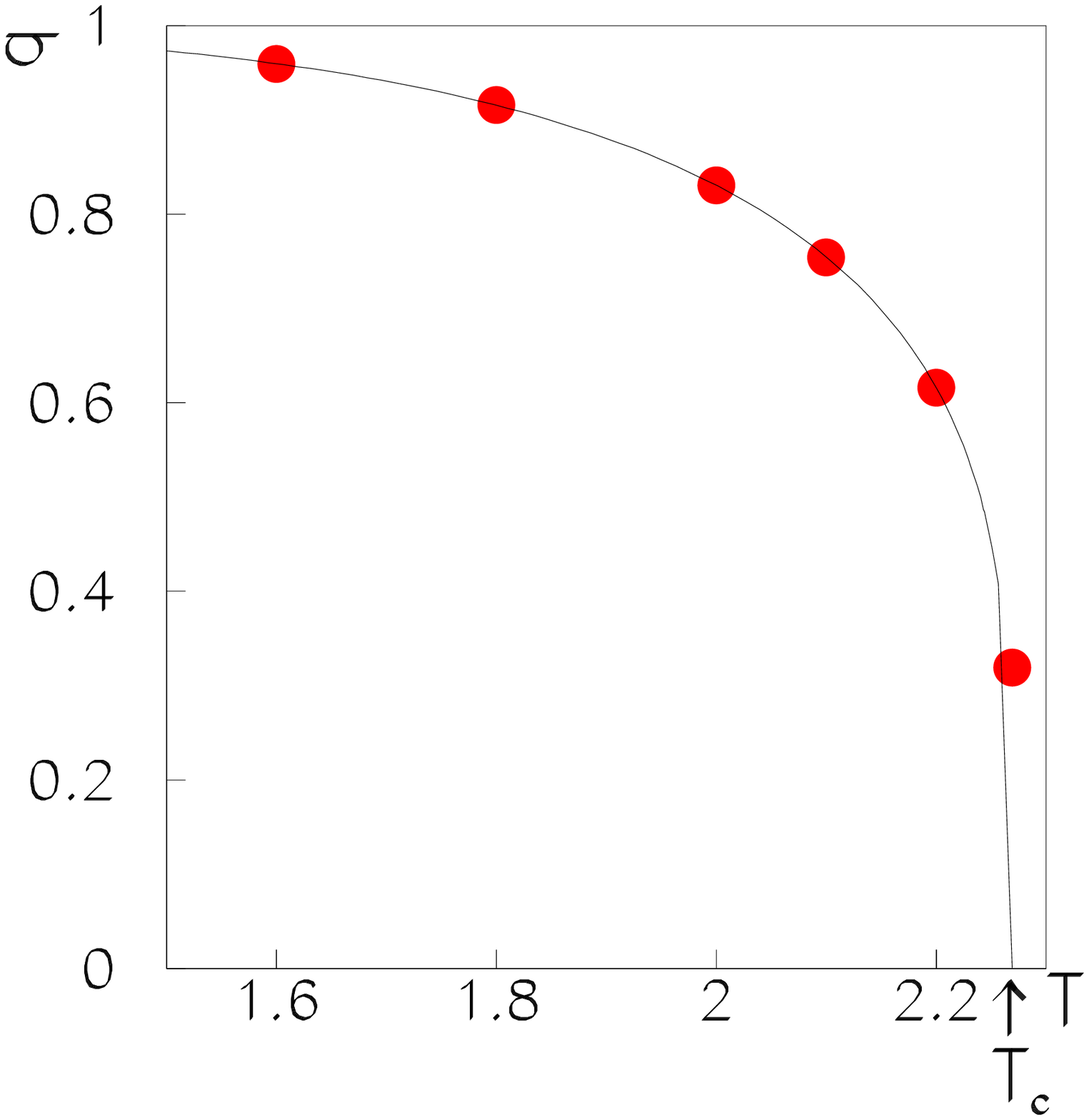}
\caption{Left panel: correlation function $C^{(p)}(r,L,T)$ against $z$
in the $2d$ Ising model with PBC for different $T$ below $T_c$ and $L=128$.
Right panel: symbols stand for the plateau height $q$ for different temperatures below $T_c$, while the continuous line is the plot of $m^2_\pm$ from Eq.  (\ref{SSB.01}).}
\label{fig:6}
\end{figure}
We may then rewrite $C^{(p)}(r,L,T)$ as the sum of two
  contributions
\be
C^{(p)}(r,L,T) = G(r,L,T) + q(T),
\label{split.02}
\ee
with $G(r,L,T)$ decaying toward zero as $r$ increases.
The origin of this additive form can be understood taking into account that
the typical configurations below $T_c$, such as
  the one depicted in the left panel of Fig.\ref{fig:conf}, 
  \begin{figure}[ht]
\centering
\includegraphics[width=5cm]{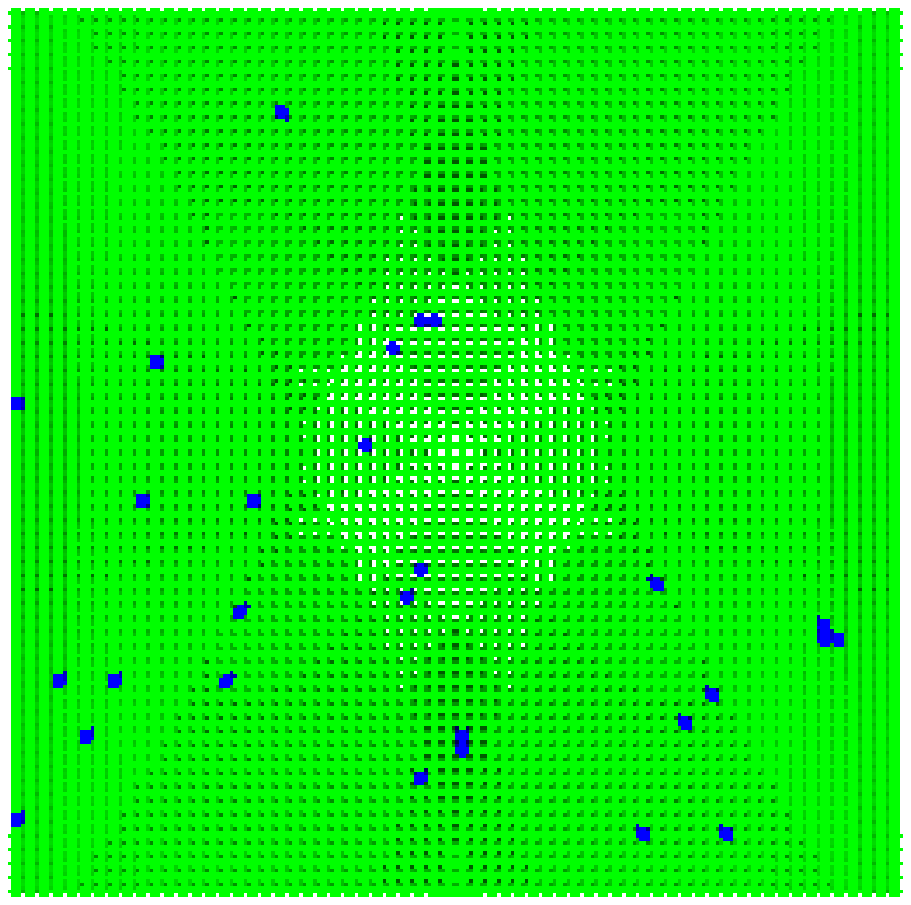}
\includegraphics[width=5cm]{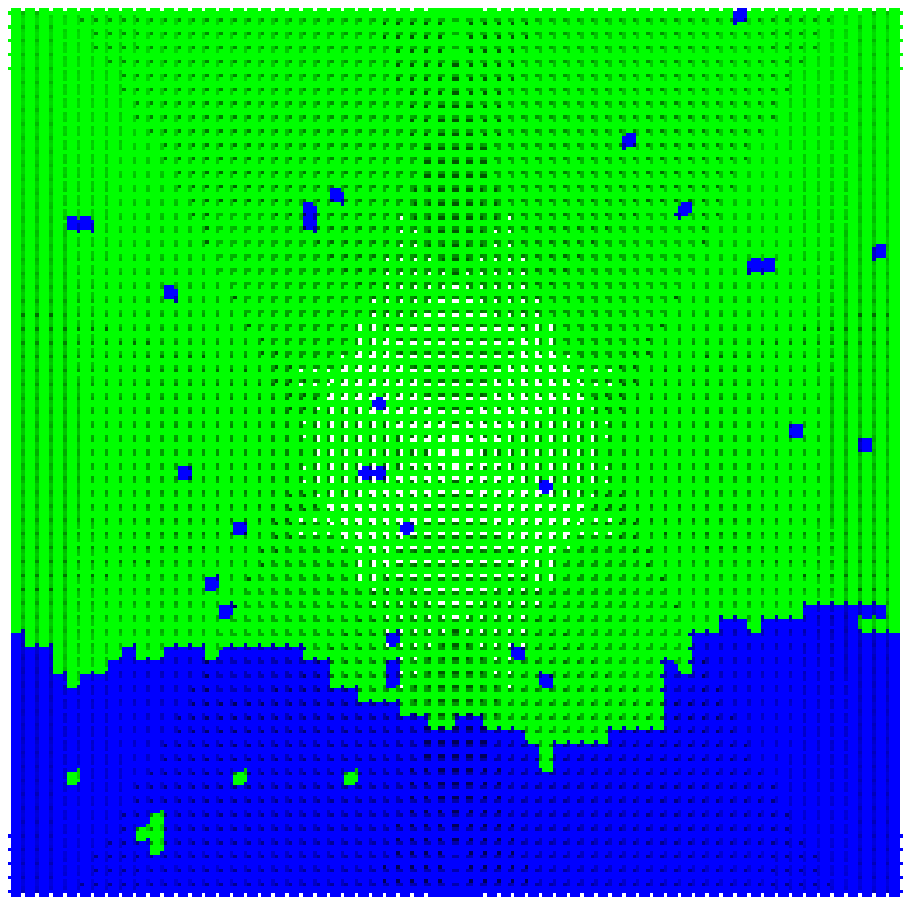}
\caption{Typical configurations with PBC (left) and APBC (right) below $T_c$.}
\label{fig:conf}
\end{figure}
display
  one ordered domain filling compactly the whole system, with small finite
  patches of reversed spins, due to thermal fluctuations. This suggests
to split the order parameter into the sum~\cite{MZ}
\be
s_i = \psi_i + \sigma, 
\label{split.01}
\ee
where $\sigma$ is a random variable which takes with probability $1/2$ the
two values $m_\pm$ and $\psi_i$ represents the thermal fluctuations
in the broken symmetry state with the signature carried by $m_\pm$. 
From the statistical independence of these two variables and the zero averages
$\langle \psi_i \rangle = \langle \sigma \rangle = 0$, one has
\be
\langle s_i s_j \rangle = \langle \psi_i\psi_j \rangle + \langle \sigma^2 \rangle,
\label{split.03}
\ee
where $\langle \sigma^2 \rangle = m^2_\pm$ is the variance of $\sigma$ and
$\langle \psi_i\psi_{i+r} \rangle$ can be identified with
the correlation function $C_\pm(r,L,T)$ in the broken symmetry pure states,
which is independent of the $\pm$ signature and obeys finite size scaling of the form 
\be
C_\pm(r,L,T) \simeq \frac{1}{r^{1/4}} g(z,\zeta).
\label{sim2.04}
\ee
Therefore, comparing Eqs.~(\ref{split.02}) and~(\ref{split.03}), we may identify
\be
G(r,L,T) = C_\pm(r,L,T), \quad  q(T) = m^2_\pm.
\label{comp.1}
\ee
Notice that the plateau contribution $q$ is independent of $L$. We emphasize that the presence of this plateau signals ergodicity breaking and, therefore, that the state below $T_c$ is not critical, contrary to what happens with APBC, as we shall see in the next
subsection.

\end{itemize}

\subsection{APBC}

In the APBC case the phenomenology is characterized by the $x,y$ anisotropy. Specifically,
$C_x^{(a)}(r,L,T)$ behaves similarly to $C^{(p)}(r,L,T)$, while $C_y^{(a)}(r,L,T)$ is qualitatively different.

\begin{itemize}

\item \underline{$T > T_c$} - As long as  $\zeta \gg 1$, the two functions
  \be
C_{x,y}^{(a)}(r,L,T) = \frac{1}{r^{1/4}}f_{x,y}^{(a)}(z,\zeta), 
\label{antip.1}
\ee
display $x,y$ isotropy
and independence from BC.
Anisotropy emerges when finite size effects become appreciable in
the $\zeta \lesssim 1$ region. Then,
the behaviors along $x$ and along $y$ separate (see left panel of Fig.\ref{fig:7})
according to a pattern reminiscent of the one in the left panel of Fig.\ref{fig:2}
for the $d=1$ case. Clearly, in the $L \to \infty$ limit the difference between the $x$
and $y$ directions disappears.
\begin{figure}[ht]
\centering
\includegraphics[width=5cm]{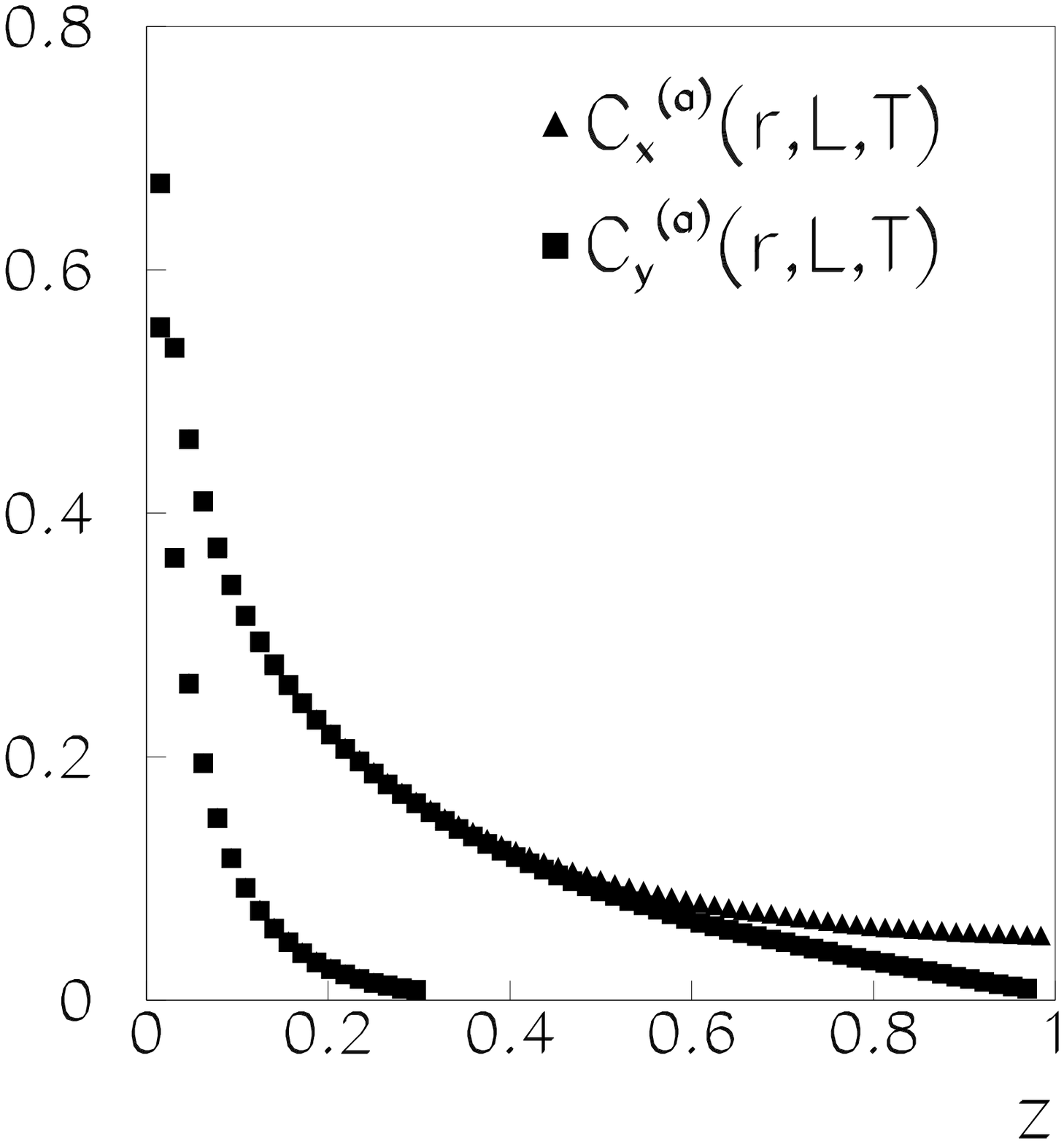}
\includegraphics[width=5cm]{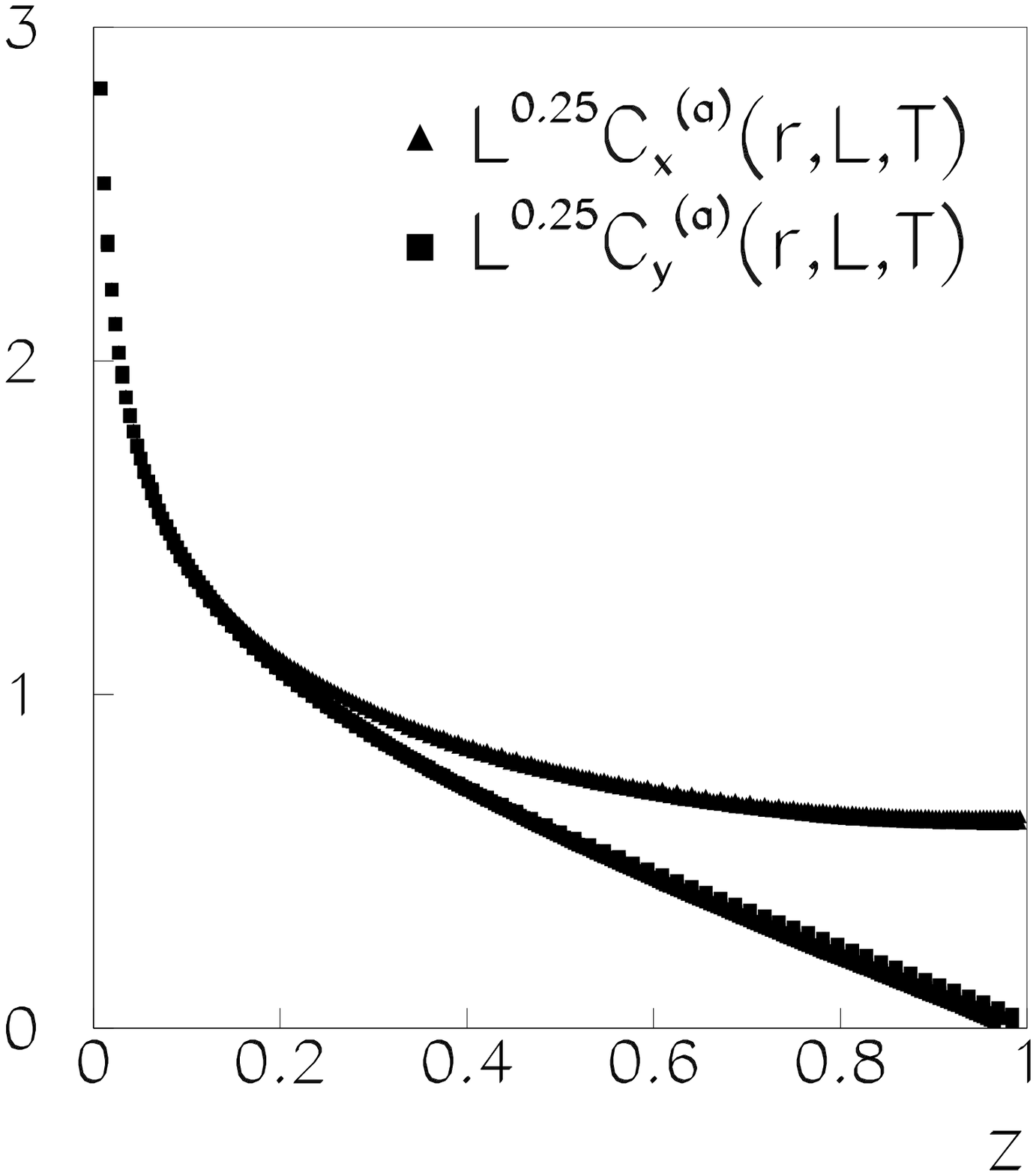}
\caption{Left panel: Correlation functions $C_x^{(a)}(r,L,T)$ (triangles) and $C_y^{(a)}(r,L,T)$ (squares)
in the $2d$ Ising model with APBC, above $T_c$ for $T=2.5$, $2.4$ (from bottom to top) and $L=128$. For $T=2.5$ the two sets of symbols superimpose.
Right panel:
$L^{0.25} C_x^{(a)}(r,L,T)$ (triangles) and $L^{0.25} C_y^{(a)}(r,L,T)$ (squares) 
(from top to bottom) in the $2d$ Ising model with APBC,
at $T_c$. Data taken with $L=~64,~128,~256,~512$ do collapse almost perfectly.
}
\label{fig:7}
\end{figure}

\item \underline{$T = T_c$} - Anisotropy of the finite size scaling at $T_c$ 
\be
C^{(a)}_{x,y}(r,L,T) = \frac{1}{r^{1/4}} f^{(a)}_{x,y}(z),
\label{sim2d.3bisbis}
\ee
is illustrated in the right panel of Fig.\ref{fig:7} by the collapse the data taken with different $L$,
when plotting $L^{0.25}C^{(a)}_{x,y}(r,L,T)$ against $z$. The anisotropy onset is at
$z \lesssim 1$.

\item \underline{$T < T_c$} -  The strong anisotropy demonstrated by the comparison of the  left and center panels of Fig.\ref{fig:8} is evidence that long range correlations
persist throughout the region $T < T_c$.
Specific features of
$C^{(a)}_{x,y}(r,L,T)$ can be accounted for by generalizing the argument previously developed
for PBC. As explained above, typical configurations (see right panel of Fig.\ref{fig:conf})
now contain two large ordered domains,
oppositely oriented and separated by a spanning domain wall, each
containing in its interior the small reversed domains due to thermal
fluctuations \cite{footnote}.
\begin{figure}[ht]
\centering
\includegraphics[width=5cm]{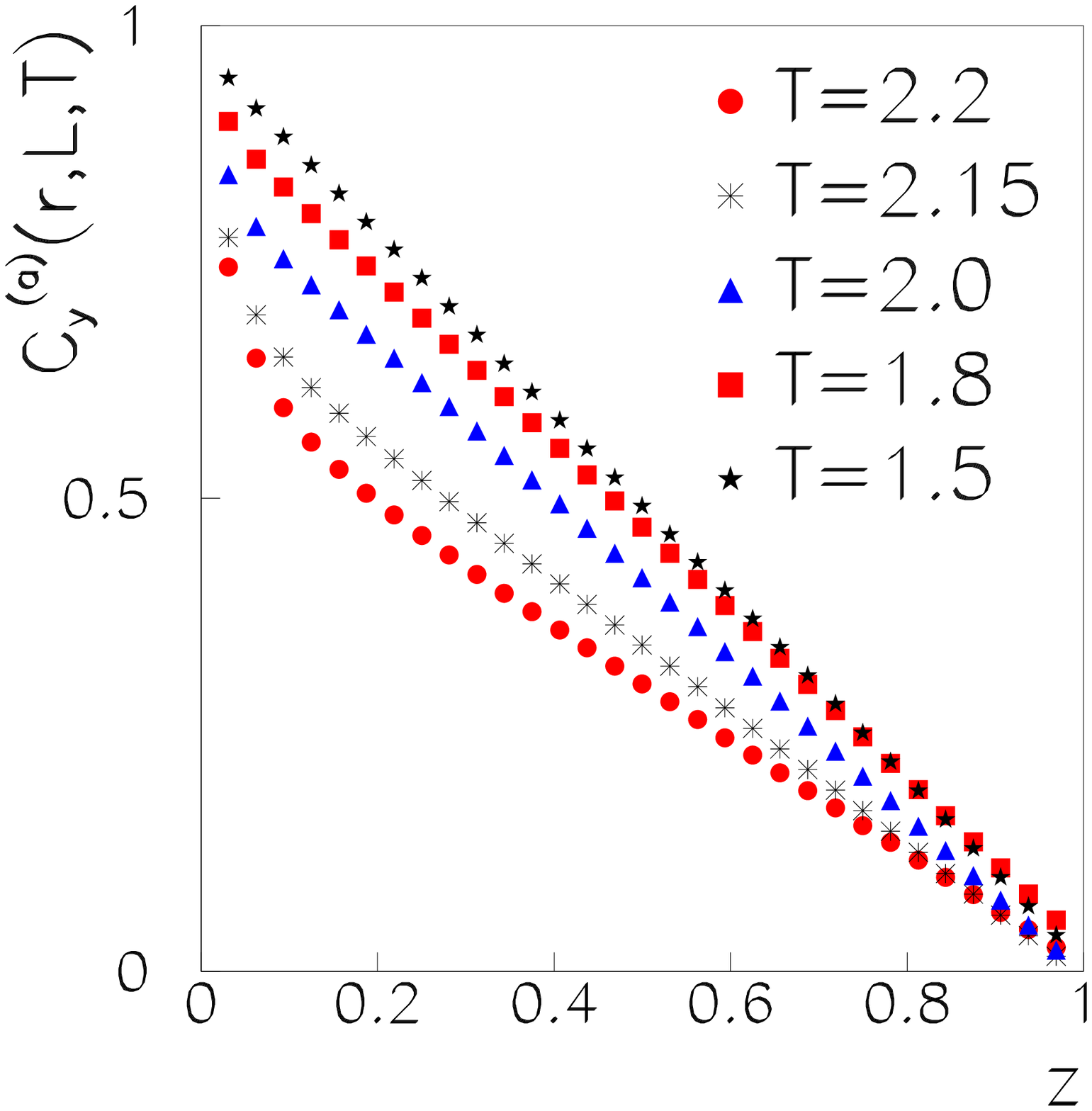}
\includegraphics[width=5cm]{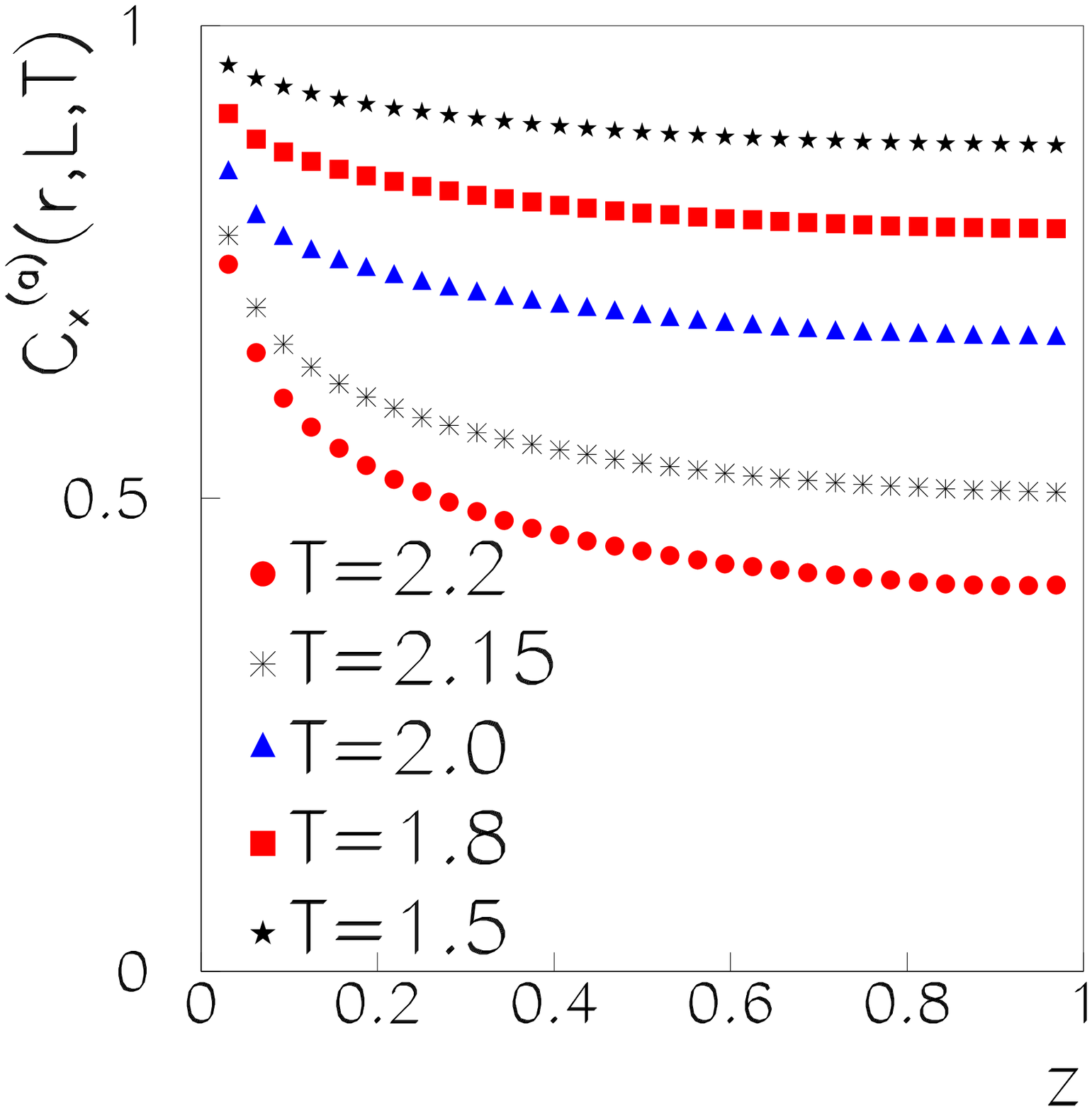}
\includegraphics[width=5cm]{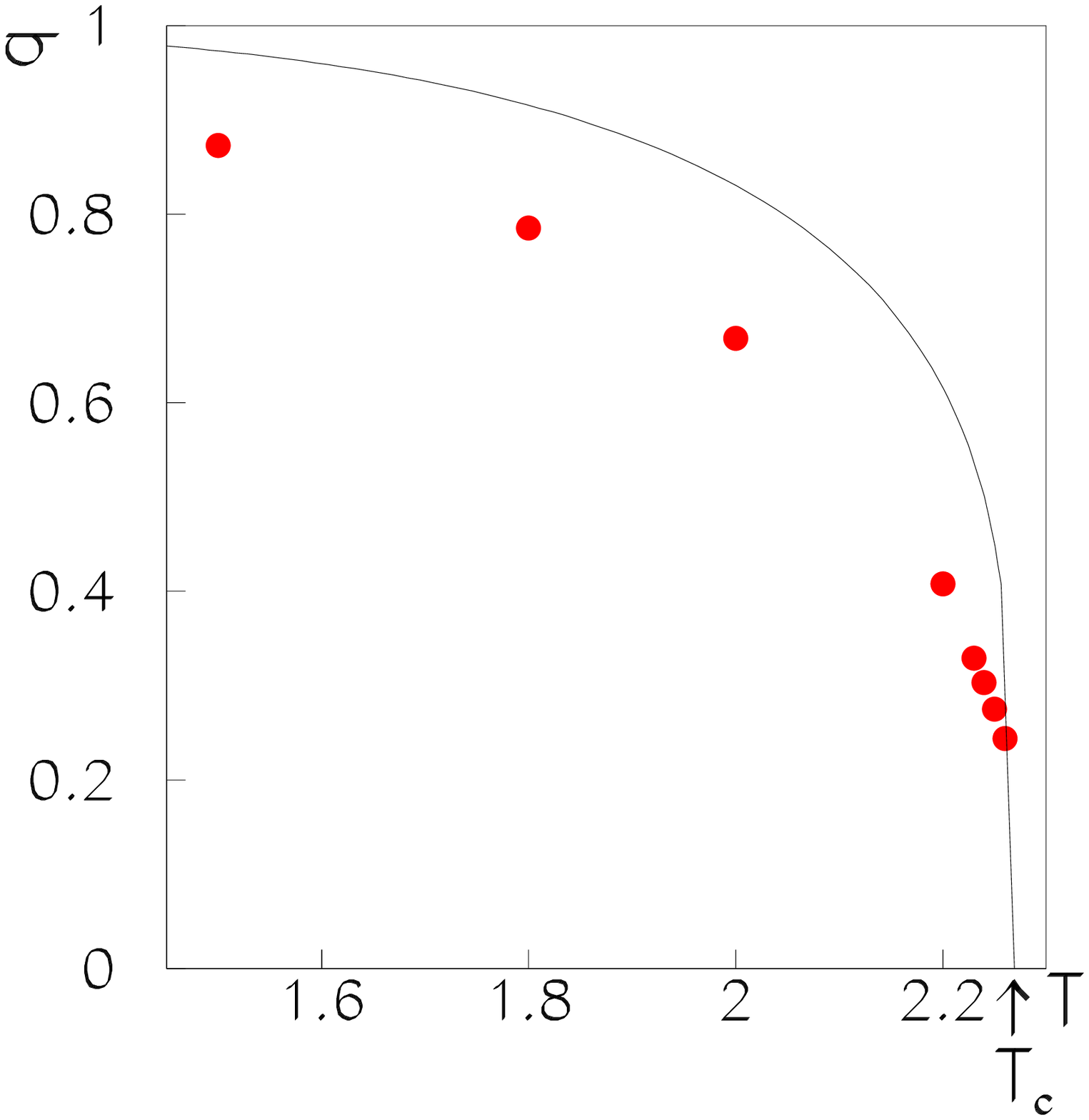}
\caption{Correlation functions $C_y^{(a)}(r,L,T)$ and $C_x^{(a)}(r,L,T)$ along the $y$ direction (left) and $x$ direction (center) with APBC, for different $T$ below $T_c$ and $L=64$. 
In the right panel the symbols stand for the plateau height from $C_x^{(a)}(r,L,T)$,
while the continous line is the plot of $m^2_\pm$ from Eq.  (\ref{SSB.01}).}
\label{fig:8}
\end{figure}
Accordingly, the split~(\ref{split.01}) of the order parameter
now must take the local form
\be
s_i = \psi_i + \sigma_i, 
\label{split.1}
\ee
where $\sigma_i = \langle s_i \rangle_{\alpha_i}$ is the average magnetization
in the broken symmetry
state with the signature $\alpha_i=\pm$ of the domain to which the site $i$ belongs,
and $\psi_i$ is the thermal fluctuations contribution.
In other words, $\sigma_i$ is a stochastic variable which flips between $m_+$ and $m_-$ as
the spanning interface crosses the site $i$.
Then, assuming statistical independence of $\psi_i$ and $\sigma_i$, and taking into account
that the average over the whole system yields
$\langle \psi_i \rangle = \langle \sigma_i\rangle =0$, one has
\be
C^{(a)}_{x,y}(r,L,T) = \langle s_is_{i+r} \rangle_{x,y} =
\underbrace{\langle \psi_i\psi_{i+r} \rangle_{x,y}}_{G(r,L,T)} + 
\underbrace{\langle \sigma_i\sigma_{i+r} \rangle_{x,y}}_{D_{x,y}(r,L,T)},
\label{split.2}
\ee
where $G(r,L,T)$ is the same thermal contribution discussed in Eq.(\ref{comp.1}) of the PBC case (and, therefore, BC-independent).
The second bulk contribution $D_{x,y}(r,L,T)$ is strongly anisotropic and,
by analogy with the arguments put forward in the $d=1$ case at $T=0$, is expected to have the structure
\begin{eqnarray}
D_{x}(r,L,T) & = & q, \\
D_{y}(r,L,T) & = & q\left (1-z \right ),
\label{D.2}
\end{eqnarray}
where the $T$ dependence is absorbed in $q$.
This is corroborated by the plots in Fig.\ref{fig:8}. Notice that the plateau height $q$ of $C^{(a)}_{x}(r,L,T)$ is somewhat lower than
$m_\pm^2$ (right panel of Fig.\ref{fig:8}), because
the interface is corrugated along the $x$ direction. The finite width of the strip occupied by the interface
induces a correction on the plateau value. 
In this connection, we expect that in the $3d$ case, where there is a finite roughening temperature $0 < T_R < T_c$, the same effect would be observed above $T_R$, where the interface is rough, but not below $T_R$, where the surface is stable. More precisely, if APBC in the $3d$ case are imposed along the $z$ direction, the plateau values of the correlation functions along the $x$ and $y$ transverse directions are expected to lie somewhat below the $m^2_\pm$ curve for $T$ chosen in between $T_R$ and $T_c$, just as in the right panel of Fig.\ref{fig:8}, while they should fall right on top of it for $T$ below $T_R$.

Apart from this, from
Eq.~(\ref{D.2}) there follows, on the basis of the considerations made at the end of
section~\ref{sec2}, that below $T_c$ the $\sigma$ degrees of freedom are critically correlated with compact clusters. In order to further elaborate on this important feature, it is instructive to consider the circularly averaged correlation function, which can be decomposed as before into the sum of two parts as a consequence of the split (\ref{split.1}) of variables
\be
C^{(a)}(r,L,T)  =G(r,L,T) + D(r,L,T).
\label{split.3}
\ee
By taking $T$ low enough to suppress the thermal contribution, the plot of 
$C^{(a)}(r,L,T)$ against $z$, for different values of $L$, in Fig.\ref{fig:11} reveals the critical behavior of the type discussed in Eq. (\ref{p.7}), characterized by  scaling and linear decay 
$D(r,L,T) \sim 1-z$.
\begin{figure}[ht]
\centering
\includegraphics[width=7.5cm]{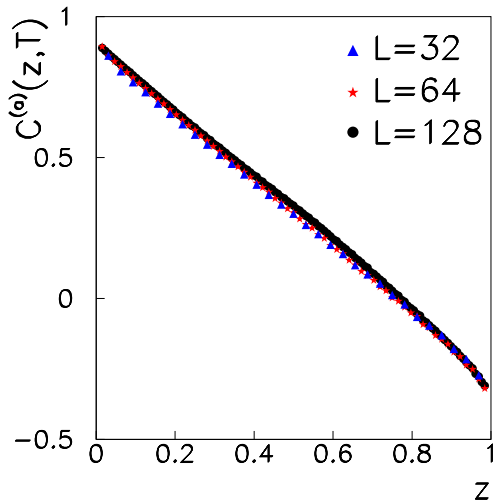}
\caption{Circularly averaged correlation function at $T=1.8$ for different $L$ values.}
\label{fig:11}
\end{figure}
This confirms that at any $T$ below $T_c$ the system with APBC is at criticality.
The picture is completed introducing critical exponents.
From the scalings of the order-parameter-like quantity $\sqrt{<m^2>}\sim L^0$ and susceptibility
$\chi = \int d\vec r \,  D(r,L,T) \sim L^d$
follow the exponent relations
$$
\beta/\nu =0 , \,\,\, \gamma/\nu=d,
$$
which, in turn, imply hyperscaling to be satisfied
\be 
2\beta +\gamma=\nu d.
\label{scaling}
\ee
From the relation $ D = d-\beta/\nu$  \cite{correlated-percolation}     
it follows that the fractal dimension is, as expected,  $D=d$, consistently with the value of the fractal dimension $D=(d+2-\eta)/2$ obtained from the value of  $\eta = 2-d$, required by the form of $D(r,L,T)$. If the results obtained in this paper can be extended to any dimensions, we expect from (\ref{scaling}) that the hyperscaling relation is always verified, implying absence of an upper critical dimension.

\end{itemize}

\section{Conclusions}
\label{sec5}

In this paper we have analysed the differences between the phase transitions occurring
in the Ising model when the thermodynamic limit is taken with PBC and APBC. In the first case
the usual SSB ferromagnetic transition is observed. In the second one there is
no ordering and no SSB. The transition consists in the onset of a regime of
critical fluctuations below $T_c$, referred to as condensation of fluctuations. 
In order to understand in what ways condensation contraposes to
the usual ferromagnetic transition, it is instructive to recall the similar dichotomy arising
when the spherical model \cite{BK} and the mean-spherical model \cite{LW} are compared in the low temperature region.
Let us organize the discussion around 
the mechanism driving the transition. In the Ising case, this can be traced back to the basic identity $\frac{1}{N}\sum_i s_i^2 = 1$ by introducing the set of variables $[m,\psi_i]$, where 
$m=\frac{1}{N}\sum_i s_i$ is the overall magnetization and the 
$[\psi_i = s_i - m]$ contain the excitations with respect to the background. Then,
the identity takes the form $m^2 + \frac{1}{N}\sum_i \psi_i^2 = 1$,
which after averaging yields
\be
\langle m^2 \rangle + \frac{1}{N}\sum_i \langle \psi_i^2 \rangle = 1.
\label{concl.1}
\ee
The above is a sum rule which must be satisfied, no matter what ensemble is used in taking the average. $T_c$ is the temperature above which the 
excitations contribution suffices to saturate the equality, while
below $T_c$ it falls short of it and the missing piece must be compensated by a finite
contribution $\langle m^2 \rangle$ coming from $m$. This is the point where the role of the statistical ensemble becomes crucial, since there is not a unique way to do it. With PBC a finite value of $\langle m^2 \rangle$ is built up by ordering. With APBC, instead, since ordering cannot take place, a finite $\langle m^2 \rangle$ comes from fluctuations condensing into the single degree of freedom $m$. Going over to the spherical models,
the transition is driven by a mechanism with an identical structure, that is a sum rule analogous to (\ref{concl.1})  originating from the spherical constraint on a continuous order parameter $\varphi(\vec x)$. In terms of Fourier components, this reads 
\be
\langle m^2 \rangle + \frac{1}{V^2}\sum_{\vec k \neq 0}\, \langle \varphi^2_{\vec k} \rangle = 1,
\label{concl.2}
\ee
where $m=\frac{1}{V}\varphi_0$ is the zero wave vector component. In the formulation of Berlin and Kac \cite{BK}, where the spherical constraint is imposed sharply, the  $\langle m^2 \rangle$ contribution,  needed
to satisfy the equality below $T_c$, comes from ordering, as in the Ising case with PBC. Instead, in the softer version of Lewis and Wannier \cite{LW}, ordering is not possible because the constraint is imposed on average,
leaving the model formally linear. Then, the required $\langle m^2 \rangle$ contribution
comes from condensation of the fluctuations of $m$ \cite{CCZ}, as in the APBC case.

Another instance of lack of ensemble equivalence, which involves ordering vs condensation, arises with the treatment of the ideal Bose gas in the canonical and grand canonical ensemble, whose analogy with the spherical models has been known for quite sometime \cite{KT}. Bose-Einstein condensation (BEC) in the canonical case
corresponds to the spontaneous breaking of gauge symmetry and to ordering. Conversely, BEC in the grand canonical framework rather fits into the condensation of fluctuations scheme \cite{EPL}. Furthermore, the discussion in the present paper gives the opportunity to comment on the use of
Griffiths inequality in Eq.~(\ref{Griffiths.1}) to establish a relation between BEC and SSB.  The inequality, properly reformulated ~\cite{Roepstorff} in
the BEC context, has been used~\cite{Lieb,Yukalov}
to argue that BEC is in fact equivalent to the spontaneous breaking of
the underlying global gauge symmetry. Keeping on using the magnetic language, the argument is based on the observation that, given the inequality,
then from
$\langle m^2 \rangle > 0$ follows $m_\pm^2 > 0$. However, such an implication is inconsequential
with regard to SSB occurrence, because Griffiths theorem, and therefore the inequality,
say nothing about the form of the probability distribution of $m$. SSB occurs only if $P(m)$
is of the bimodal form as sketched in Fig.\ref{fig:1}.
The results presented above do clarify the issue by showing, in the transparent
context of the Ising model, that the inequality may well be satisfied, as in Eq.~(\ref{ens.5}) above,
and yet the transition to occur without SSB.

As a final comment, we point out that the critical state below $T_c$ arising
with APBC is of considerable interest also in the framework of the phase ordering
process following the quench from above to below $T_c$.
Recall that phase ordering~\cite{Bray,Puri,MZ} is the relaxation process taking place after the sudden
temperature drop below $T_c$
of a system initially equilibrated above $T_c$, {\it with free or PBC}. Then, if the thermodynamic
limit is taken beforehand, the system remains permanently out of equilibrium, exhibiting
slow relaxation characterized by dynamical scaling and aging~\cite{MZ}.
Now, even though equilibrium is never achieved, yet the sequence
$\lim_{t \to \infty} \lim_{V \to \infty}\langle \cdot \rangle_{V,t}$
yields a well defined limit which, and here is the
point, shares the properties of the equilibrium state prepared with APBC,
even though, we emphasize, the evolution takes place with free or periodic BC.
Specifically, the putative equilibrium state to be reached by the never ending relaxation process of phase ordering is critical with compact correlated domains, just as in the APBC equilibrium state.
The investigation of this important connection is the object of work in preparation.

\section*{Acknowledgments}
We thank an unknown referee for comments and suggestions which led to an
improvement in the presentation of the paper.
A.C. and A.F. acknowledge financial support from the CNR-NTU joint laboratory {\it Amorphous materials for energy harvesting applications}.


\begin{thebibliography}{99}


\bibitem{Landau}
  L. D. Landua and E. M. Lifshitz, {\it Statistical Physics}, 3d Edition Pergamon Press (1980).

\bibitem{Palmer}
  R. G. Palmer, Adv. Phys. {\bf 31}, 669 (1982).

\bibitem{van Enter}
  A. C. D. van Enter and J. L. van Hemmen, Phys. Rev. A {\bf 29}, 355 (1984).

\bibitem{Griffiths}
  R. B. Griffiths,  Phys. Rev. {\bf 152}, 240 (1966).

\bibitem{Gallavotti}
G. Gallavotti, Riv. Nuovo Cimento {\bf 2}, 133 (1972);
{\it Statistical Mechanics A Short Treatise}, Springer-Verlag Berlin Heidelberg 1999.

\bibitem{Delfino}
  G. Delfino, W. Selke and A. Squarcino, arXiv:1803.04759v1 [cond-math.stat-mech].

\bibitem{Hasenbusch}
M. Hasenbusch and S. Meyer, Phys. Rev. Lett. {\bf 66}, 530 (1991).

\bibitem{Antal}
  T. Antal, M. Droz and Z. R\'acz, J. Phys. A: Math. Gen. {\bf 37}, 1465 (2004).


\bibitem{BCKM}
  J. P. Bouchaud, L. F. Cugliandolo, J. Kurchan and M. Mezard, {\it Out of equilibrium dynamics in
    spin glasses and other glassy systems}, in {\it Spin Glasses and Random Fields} A. P. Young ed.,
  Singapore: World Scientific 1997.


 \bibitem{MZ}
M. Zannetti, in {\it Kinetics of Phase Transitions}, S. Puri and V. Wadahawan Eds., CRC Press 2009.
  

\bibitem{correlated-percolation} A. Coniglio and A. Fierro (2009), {\it Correlated Percolation},
in R. A. Meyers (Ed.) Encyclopedia of
Complexity and Systems Science, Part 3, pp 1596-1615; arXiv:1609.04160.

\bibitem{McCoy}
B. M. McCoy and T. T. Wu, {\it The two dimensional Ising Model}, Harvard University Press,
Cambridge, Massachusetts 1973.


\bibitem{Bruce}
  A. D. Bruce, J. Phys. C: Solid State Physics {\bf 14}, 3667 (1981);
  J. Phys. A: Math. Gen. {\bf 18} L873 (1985).

\bibitem{Binder}
  K. Binder, Z. Phys. {\bf 43}, 119 (1981).
  
 \bibitem{footnote}
The presence of just one spanning interface is in agreement with the theorem quoted in section 11 of G. Gallavotti, Riv. Nuovo Cimento {\bf 2}, 133 (1972).
  

\bibitem{BK}
   T. H. Berlin and M. Kac, Phys. rev. {\bf 86}, 821 (1952).

 \bibitem{LW}
   H. W. Lewis and G. H. Wannier, Phys. Rev. {\bf 88}, 682 (1952) and
   Phys. Rev. {\bf 90}, 1131E (1953).

 \bibitem{CCZ}
   C. Castellano, F. Corberi and M. Zannetti, Phys. Rev. E {\bf 56}, 4973 (1997);
   N. Fusco and M. Zannetti, Phys. Rev. E {\bf 66}, 066113 (2002).
   
   \bibitem{KT}
   M. Kac and C. J. Thompson, J. Math. Phys. {\bf 18}, 1650 (1977).
   
   \bibitem{EPL}
   M. Zannetti, EPL {\bf 111}, 20004 (2015).

\bibitem{Roepstorff}
G. Roepstorff, J. Stat. Phys. {\bf 18}, 191 (1978).

\bibitem{Lieb}
  E. H. Lieb, R. Seiringer and  J. Yngvason,  Phys. Rev. Lett. {\bf 94}, 080401 (2005).


  \bibitem{Yukalov}
    V. I. Yukalov, Laser Phys. Lett. {\bf 4}, 632 (2007).


\bibitem{Bray}
  A. J. Bray, Adv. Phys. {\bf 43}, 357 (1994).


\bibitem{Puri}
   S. Puri, in {\it Kinetics of Phase Transitions}, S. Puri and V. Wadahawan Eds., CRC Press 2009.


 
    






















\end{thebibliography}
\end{document}